\documentclass[12pt]{article}

\usepackage{amsmath, amssymb, slashed}
\usepackage[dvips]{graphicx}
\usepackage{epsf}
\usepackage{color}
\usepackage{subfigure}

\begin{document}
\def\a{\alpha}
\def\b{\beta}
\def\c{\varepsilon}
\def\d{\delta}
\def\e{\epsilon}
\def\f{\phi}
\def\g{\gamma}
\def\h{\theta}
\def\k{\kappa}
\def\l{\lambda}
\def\m{\mu}
\def\n{\nu}
\def\p{\psi}
\def\q{\partial}
\def\r{\rho}
\def\s{\sigma}
\def\t{\tau}
\def\u{\upsilon}
\def\v{\varphi}
\def\w{\omega}
\def\x{\xi}
\def\y{\eta}
\def\z{\zeta}
\def\D{{\mit \Delta}}
\def\G{\Gamma}
\def\H{\Theta}
\def\L{\Lambda}
\def\F{\Phi}
\def\P{\Psi}

\def\S{\Sigma}

\def\o{\over}
\def\beq{\begin{eqnarray}}
\def\eeq{\end{eqnarray}}
\newcommand{\gsim}{ \mathop{}_{\textstyle \sim}^{\textstyle >} }
\newcommand{\lsim}{ \mathop{}_{\textstyle \sim}^{\textstyle <} }
\newcommand{\vev}[1]{ \left\langle {#1} \right\rangle }
\newcommand{\bra}[1]{ \langle {#1} | }
\newcommand{\ket}[1]{ | {#1} \rangle }
\newcommand{\EV}{ {\rm eV} }
\newcommand{\KEV}{ {\rm keV} }
\newcommand{\MEV}{ {\rm MeV} }
\newcommand{\GEV}{ {\rm GeV} }
\newcommand{\TEV}{ {\rm TeV} }
\def\diag{\mathop{\rm diag}\nolimits}
\def\Spin{\mathop{\rm Spin}}
\def\SO{\mathop{\rm SO}}
\def\O{\mathop{\rm O}}
\def\SU{\mathop{\rm SU}}
\def\U{\mathop{\rm U}}
\def\Sp{\mathop{\rm Sp}}
\def\SL{\mathop{\rm SL}}
\def\tr{\mathop{\rm tr}}

\def\IJMP{Int.~J.~Mod.~Phys. }
\def\MPL{Mod.~Phys.~Lett. }
\def\NP{Nucl.~Phys. }
\def\PL{Phys.~Lett. }
\def\PR{Phys.~Rev. }
\def\PRL{Phys.~Rev.~Lett. }
\def\PTP{Prog.~Theor.~Phys. }
\def\ZP{Z.~Phys. }

\begin{titlepage}
\begin{center}

\hfill IPMU-12-0178\\
\hfill ICRR-report-628-2012-17\\
\hfill \today

\vspace{1.5cm}
{\large\bf 
Imprints of Non-theramal Wino Dark Matter 
\\
on Small-Scale Structure
}
\vskip 1.2cm
{ Masahiro Ibe}$^{(a,b)}$,
{ Ayuki Kamada}$^{(b)}$
and
{ Shigeki Matsumoto}$^{(b)}$\\

\vskip 0.4cm
{\it
$^{(a)}${\it ICRR, University of Tokyo, Kashiwa, 277-8583, Japan }\\
$^{(b)}${Kavli IPMU, University of Tokyo, Kashiwa, 277-8583, Japan} 
}

\vskip 1.5cm

\abstract{
We study how ``warm'' the wino dark matter is when it is non-thermally produced by the decays of the gravitino in the early Universe.
We clarify the energy distribution of the wino at the decay of the gravitino and the energy loss process after their production.
By solving the Boltzmann equation, we show that a sizable fraction of the wino dark matter can be ``warm" for the wino mass $m_{\tilde w}\simeq 100-500\,{\rm GeV}$.
The ``warmness'' of the wino dark matter leaves imprints on the matter power spectra and may provide further insights on the origin of dark matter via the future $21\,{\rm cm}$ line survey.
Our calculations can be applied to other non-thermal wino production scenarios such as the wino dark matter produced by the decay of the moduli fields. 
}

\end{center}
\end{titlepage}
\setcounter{footnote}{0}

\section{Introduction}
\label{sec: introduction}

The existence of dark matter in the Universe has been 
established by numerous cosmological and astrophysical observations
on a wide range of scales. 
Its nature, however, has remained unknown for almost eighty years since its first postulation, and hence, 
the identification of dark matter is arguably the most important challenge 
in cosmology, astrophysics, and particle physics\,\cite{darkmatter}.

Although we know little of the origin of dark matter,
we are (almost) certain that dark matter is not a part of the standard model.
Therefore, it is highly motivated to relate the identity of dark matter 
with physics beyond the standard model\,\cite{Murayama:2007ek}.
In particular, the supersymmetric standard model\,\cite{Fayet} is one of the most 
viable candidates of new physics which contains a good candidate for dark matter, i.e.
the lightest supersymmetric particle (LSP).
Supersymmetry (SUSY) is attractive since it allows
the vast separation of low energy scales from high energy scales
such as the Planck scale or the scale of the Grand Unified Theory (GUT).
The supersymmetric standard model is also supported by the precise unification 
of three gauge coupling constants of the standard model at the GUT scale.

In this paper, we consider the wino LSP dark matter scenario
where the relic density of the wino is provided by the late time 
decays of the heavy gravitino\,\cite{Gherghetta:1999sw, hep-ph/9906527,Ibe:2004tg}.
Due to the large mass hierarchy between the gravitino and the wino, 
the produced wino is more energetic than the thermal background.
Thus, the wino LSP can be ``warmer'' than the conventional cold dark matter
and leaves imprints on the small-scale structure if it does not lose its energy 
via the scattering processes with the thermal background.
As we will show, a sizable fraction of the wino dark matter can be ``warm" for the wino mass $m_{\tilde w}\simeq 100-500\,{\rm GeV}$.
The imprints on the matter power spectra may provide further insights on the origin of dark matter via the future $21{\rm cm}$ line survey\,\cite{21cmcosmology}.

Here, we mention that the wino LSP scenario is now highly motivated
after the discovery of a Higgs-like particle with a mass around $125$\,GeV
at the LHC experiments\,\cite{ATLAS,CMS}.
As is well known, the lightest Higgs boson mass is strongly interrelated to 
the sfermion masses\,\cite{Higgs,TU-363} in the minimal supersymmetric standard model (MSSM).
The observed Higgs boson mass around $125$\,GeV, then, suggests
that the sfermion masses are in the tens to hundreds TeV range\,\cite{TU-363}.
In the simplest supersymmetry breaking mediation mechanism, i.e. gravity mediation\,\cite{SUGRA},
such heavy sfermions are realized when the gravitino mass  is 
in the tens to hundreds TeV range.
The gaugino masses are, on the other hand, in about a TeV range 
even for such a heavy gravitino when they are generated radiatively, which
is expected when there is no singlet supersymmetry breaking field, i.e. the Polonyi field.
The Polonyi field is disfavored from cosmology,
since it causes the so-called Polonyi problem\,\cite{Polonyi,hep-ph/0605252}.
Finally, the higgsino can be as heavy as the gravitino 
in ``pure gravity mediation model''\,\cite{Ibe:2006de,Ibe:2011aa, Ibe:2012hu},
where the so-called $\mu$-term of the order of the gravitino mass is generated 
without use of the Polonyi field.
Therefore, by assuming the simplest model based on the supergravity without the Polonyi field,
the Higgs boson mass around $125$\,GeV naturally leads to the wino LSP scenario 
in the hundreds GeV range (see Refs.\,\cite{Ibe:2011aa, Ibe:2012hu} for details).

The organization of the paper is as follows.
In section\,\ref{sec:WinoLSP}, we summarize the wino LSP scenario
mainly assuming the pure gravity mediation model, although
our discussion can be applied for generic wino LSP scenarios.
In section\,\ref{sec:SmallScale}, we discuss the imprints on the 
small-scale structure of the wino dark matter
produced non-thermally by the decays of the gravitino.
The final section is devoted to conclusions.

\section{The Wino LSP Scenario}
\label{sec:WinoLSP}

\subsection{Summary of the model}
First, let us summarize the  the wino LSP scenario.
To be specific, we base our arguments on the pure gravity mediation
model\,\cite{Ibe:2006de,Ibe:2011aa, Ibe:2012hu}, although
the most of the following discussions can be applied to generic wino LSP scenarios
with the heavy sfermions and higgsinos.%
\footnote{
The pure gravity mediation model summarized below has many common features 
with the PeV-scale Supersymmetry\,\cite{Wells:2004di},
the $G_2$--MSSM\,\cite{Acharya:2007rc},
the Spread Supersymmetry\,\cite{Hall:2011jd},
and the model with strong moduli stabilization\,\cite{Dudas:2012wi}.
The model also has a certain resemblance to the Split Supersymmetry\,\cite{hep-th/0405159,hep-ph/0406088,hep-ph/0409232}, where the higgsino is assumed to be in the TeV scale. (See also \cite{Bose:2012gq} for discussions on the origin of the $\mu$-term.)
}

In our setup, scalar fields in the MSSM obtain soft supersymmetry breaking masses 
via tree-level interactions in supergravity. 
With a generic K\"ahler potential, all the supersymmetry breaking masses of the scalar bosons are expected to be of the order of the gravitino mass, $m_{3/2}$.
For  simplicity,  we assume that all the sfermions have the gravitino mass,
\begin{eqnarray}
m_{\rm sfermion}^2 \simeq m_{3/2}^2\ ,
\end{eqnarray}
in the following discussions, 
although the details of the sfermion mass spectra do not change the discussions significantly. 
To account for a recently observed Higgs boson mass around $125$\,GeV,
we take $m_{3/2}$ in tens to hundreds TeV range (see Refs.\,\cite{Ibe:2011aa, Ibe:2012hu} for example).

The higgsino mass, it is also generated through tree-level interactions 
in the pure gravity mediation model using the mechanism in Ref.\,\cite{Inoue:1991rk}. 
As a result, the higgsino masses as well as the heavy Higgs boson masses
are again of the order of the gravitino mass.
It should be noted that 
a linear combination of the Higgs doublet bosons, 
\begin{eqnarray}
h \simeq \sin \beta\, H_u - \cos\beta\, H_d^*\ ,
\end{eqnarray}
should be light so that it can play a role of the standard model Higgs boson, 
which requires fine-tuning between the Higgs mass parameters to some extent.  

The gaugino masses are, on the other hand, dominated by loop suppressed contributions; 
the anomaly mediated contributions~\cite{Giudice:1998xp, Randall:1998uk,hep-ph/9205227}, 
and the contributions from the heavy 
higgsino threshold effect\,\cite{Gherghetta:1999sw,Giudice:1998xp, Ibe:2006de}. 
For $m_{3/2}=O(10-100)\,{\rm TeV}$, the resultant physical gaugino masses are given by,
\begin{eqnarray}
\label{eq:gluino}
m_{\tilde g} &\simeq&
2.5\times (1 - 0.13 \, \delta_{32} - 0.04 \, \delta_{\rm SUSY})
\times 10^{-2} \, m_{3/2},
\\ \label{eq:wino}
m_{\tilde w} &\simeq&
3.0\times(1 - 0.04 \, \delta_{32} + 0.02 \, \delta_{\rm SUSY})
\times 10^{-3} \, (m_{3/2} + L),
\\ \label{eq:bino}
m_{\tilde b} &\simeq&
9.6\times(1 + 0.01 \, \delta_{\rm SUSY})
\times 10^{-3} \, (m_{3/2} + L/11),
\end{eqnarray}
where the subscripts $\tilde g$, $\tilde w$ and $\tilde b$ denote the gluino, the wino and the bino, respectively.
Here, $\delta_{\rm SUSY} = \log[m_{\rm sfermion}/100 \,{\rm TeV}]$, 
$\delta_{32} = \log[m_{3/2}/100 \, {\rm TeV}]$ for the gluino mass, 
and $\delta_{32} = \log[(m_{3/2} + L) /100 \,{\rm TeV}]$ for the wino mass. 
The terms proportional to $m_{3/2}$ in above formulae represent the anomaly mediated contributions, while those proportional to $L$ are the higgsino threshold contributions. 
As discussed in Ref.~\cite{Ibe:2011aa, Ibe:2012hu}, the parameter $L$ is 
of the order of the gravitino mass in the 
pure gravity mediation model, while 
$L$ is expected to be small in the conventional anomaly mediation
models.
The above formulae show that the wino is the LSP for $|L| \lesssim 3 m_{3/2}$.

Before closing this section, we here summarize the important properties of the wino LSP
which are relevant for the later discussion.
First, the mass of the neutral component (the neutral wino LSP $\tilde w^0$) is degenerated with 
the one of its charged component (the charged wino $\tilde w^\pm$) 
due to the approximate custodial symmetry. 
The dominant mass splitting between the charged and the neutral winos 
comes from one-loop gauge boson 
contributions\,\cite{Feng:1999fu}, which is given by
\begin{eqnarray}
\label{eq:deltaM}
{\mit \Delta} m_{\tilde w}  = m_{\tilde w^\pm}- m_{\tilde w^0} =
\frac{g_2^2}{16\pi^2} m_{\tilde w}
\left[ f(r_W) - \cos^2 \theta_W f(r_Z) - \sin^2 \theta_W f(0) \right],
\end{eqnarray}
where $f(r)= \int^1_0 dx (2 + 2 x^2) \ln[x^2 + (1 - x)r^2]$ and $r_{W,Z}=m_{W,Z}/m_{\tilde w}$. For the wino mass in the hundreds GeV range, the splitting is ${\mit \Delta} m_{\tilde w} \simeq 160-170$ MeV. 
Due to the mass degeneracy, 
the dominant decay mode of the charged wino is the one into a neutral wino and a charged pion, 
${\tilde w}^\pm \to {\tilde w}^0 + \pi^\pm$. 
Therefore, the charged wino has a rather long lifetime,
\begin{eqnarray}
\tau_{w^\pm} \simeq 1.2
\times 10^{-10} {\rm sec}
\left( \frac{160 \, \rm MeV}{\mit \Delta m_{\tilde w}} \right)^3
\left(1 - \frac{m_\pi^2}{{\mit \Delta}m_{\tilde w}^2} \right)^{-1/2} .
\end{eqnarray}
As we will see in the following section, the charged wino produced 
by the gravitino decays loses most of its energy before it decays 
due to this long lifetime.

\subsection{Relic density of the wino LSP}
Due to its large annihilation cross section, the thermal relic number density of the wino LSP is suppressed, 
and hence, the wino mass consistent with the observed dark matter density is rather high.
In fact, the thermal relic density,  $\Omega^{(TH)} h^2$,
in Ref.\,\cite{hep-ph/0610249}
shows that the observed dark matter density
$\Omega^{(TH)} h^2\simeq 0.11$\,\cite{Komatsu:2010fb} 
is obtained for  $m_{\tilde w}\simeq2.7$\,TeV,
while it is quickly decreasing for the lighter wino.

In the present scenario, we have another source of the wino LSP, 
the late time decays of the gravitino.
After inflation, the gravitino is copiously produced from the thermal background
and the resultant yield before its decay is roughly given by\,\cite{Bolz:2000fu},
\begin{eqnarray}
Y_{3/2} \simeq 1.9 \times 10^{-12}\left( \frac{T_R}{10^{10}\,{\rm GeV}}\right)\ ,
\end{eqnarray}
where $T_R$ is the reheating temperature of the Universe after inflation.
The produced gravitino eventually decays into gauginos at a late time
with a decay rate,
\begin{eqnarray}
\label{eq:gravitino}
  \Gamma_{3/2} &\simeq& \frac{1}{32\pi}(8+3 +1) \frac{m_{3/2}^3}{M_{\rm Pl}^2}\ , \cr
  &\simeq& 1.8\times 10^{-23}\,{\rm GeV}\times \left( \frac{m_{3/2}}{100\,\rm TeV}\right)^3\ ,
\end{eqnarray}
which corresponds to the cosmic temperature around
\begin{eqnarray}
 T_{d}\simeq 3.8\,{\rm MeV} \times  \left(\frac{m_{3/2}}{100\,{\rm TeV}}\right)^{3/2}\ .
\end{eqnarray}
Here,  $M_{\rm Pl} \simeq 2.44\times 10^{18}$\,GeV is the reduced Planck scale
and we have defined the decay temperature by,
\begin{eqnarray}
 T_d = \left( \frac{10}{\pi^2 g_*} M_{\rm Pl}^2 \Gamma_{3/2}^2  \right)^{1/4} \ .
\end{eqnarray}
Here, $g_*\simeq 10.75$ is the effective number of the massless degrees of freedom 
at around the decay time.
In the above decay width in Eq.\,(\ref{eq:gravitino}), 
we have assumed that the gravitino decays into gauginos and 
all the other modes are closed.
If there are additional decay modes into the squarks, 
the gravitino decay width becomes slightly larger, 
although it is in the same order of magnitude.
It should be noted that the decay temperature of the gravitino is high
enough not to spoil the success of the Big-Bang Nucleosynthesis
for $m_{3/2}\gtrsim 30$\,TeV\,\cite{kkm}.

As the result of the late time decays of the gravitino, the wino LSP is non-thermally
produced at around $T_d$ which is lower than the
freeze-out temperature of the wino LSP, 
and the resultant relic density is given by,
\begin{eqnarray}
 \Omega^{(NT)} h^2 \simeq 0.16 \times \left(  \frac{m_{\tilde w}}{300\,{\rm GeV}} \right)
 \left(  \frac{T_R}{10^{10}\,{\rm GeV}} \right)\ .
\end{eqnarray}
Altogether, the total relic density of the wino LSP is given by,
\begin{eqnarray}
\Omega h^2 = \Omega^{(TH)}h^2+  \Omega^{(NT)} h^2\ .
\end{eqnarray}
Thus, for $m_{\tilde w} \ll 2.7$\,TeV, the observed dark matter density 
can be explained by the non-thermally produced wino for
\begin{eqnarray}
\label{eq:TR}
T_R \simeq 7 \times 10^{9}\,{\rm GeV}
\times
\left(\frac{300\,{\rm GeV}}{m_{\tilde w}}\right)
\left(\frac{\Omega h^2}{0.11}\right)\ .
\end{eqnarray}
Interestingly, the required reheating temperature is consistent with
the successful leptogenesis\,\cite{leptogenesis}, $T_R\gtrsim 2\times 10^9$\,GeV, 
for $m_{\tilde w}\lesssim 1$\,TeV.
In the following analysis, we focus on the wino mass below a TeV
where the non-thermally produced wino is the dominant component of dark matter.

Several comments are in order.
First, it should be noted that the entropy produced by
the decay of the gravitino is negligible since the 
energy density of the gravitino at the decay time is subdominant.%
\footnote{In the non-thermal wino production scenario by the decay
of moduli fields, a large amount of entropy 
is produced, and hence, the baryon asymmetry is highly diluted.
Therefore, those scenarios require baryogenesis below the decay temperature
of the moduli fields.
}
Second, it should be also noted that the annihilation of the wino after the non-thermal
production is negligible, since the yield of the non-thermally produced wino 
is small enough, i.e.
\begin{eqnarray}
 Y_{\tilde w}^{(NT)} = Y_{3/2} \ll \sqrt{\frac{45}{8\pi^2 g_*}} \frac{1}{\langle{\sigma v} \rangle M_{\rm Pl} T_d}
 \simeq 2.9 \times 10^{-10} \times 
 \left(\frac{10^{-24}{\rm cm^3/s}}{\langle{\sigma v}\rangle }\right)
\left(\frac{4\,{\rm MeV}}{T_d}\right)\ .
\end{eqnarray}

\subsection{Collider/Indirect dark matter search constraints}

\begin{figure}[t]
\begin{center}
\includegraphics[width=0.45\linewidth]{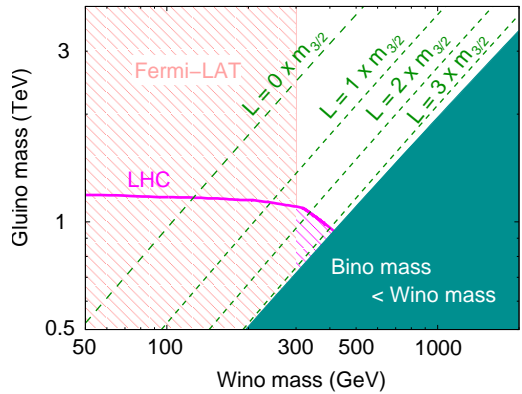}
\caption{\sl \small Constraints on the gluino and the wino masses in the pure gravity mediation model. Limits on the gluino mass obtained from the LHC experiment (8 TeV \& 6 fb$^{-1}$ data) and on the wino mass from the Fermi-LAT experiment (observation of $\gamma$-rays from dwarf spheroidal galaxies) are shown as hatched regions. The shaded region is not favored because the LSP is the bino which leads to the overclosure of the Universe.}
\label{fig: limits}
\end{center}
\end{figure}

The gluino and the wino masses in the pure gravity mediation model are constrained by the collider experiments and indirect dark matter searches. Currently, the most severe limit on the gluino mass is obtained from an inclusive SUSY search at the LHC experiment with use of multi-jets plus missing transverse energy events. According to the search, the gluino mass is constrained to be $m_{\tilde g} > 1.2\,(1.0)\,{\rm TeV}$ for the wino mass of $100\,(500)\,{\rm GeV}$~\cite{ATLAS-CONF-2012-109}, as shown in figure~\ref{fig: limits}. In near future, the limit will be increased up to about $2\,{\rm TeV}$ when the LHC experiment succeeds to accumulate $300\,{\rm fb}^{-1}$ data at the $14\,{\rm TeV}$ running. Furthermore, with utilizing disappearing charged tracks caused by long-lived charginos, the limit will be increased up to about $2.3\,{\rm TeV}$ if standard model backgrounds against the signal are efficiently reduced~\cite{Bhattacherjee:2012ed}.

On the other hand, the most severe limit on the wino mass is obtained from the indirect detection measurement of dark matter using gamma-rays from dwarf spheroidal galaxies (Milky-way satellites) at the Fermi-LAT experiment~\cite{Ackermann:2011wa}. At present, the limit is $m_{\tilde w} > 300\,{\rm GeV}$, which is also shown in figure~\ref{fig: limits}. Very recently, it has also been suggested that the observation of gamma-rays from our galactic center gives a more stringent limit on the annihilation cross section of dark matter with use of the Fermi-LAT data~\cite{Hooper:2012sr}. According to this analysis, the wino mass will be constrained as $m_{\tilde w} > 500\,{\rm GeV}$. In future, the indirect detection measurement of dark matter using cosmic-ray anti-protons at the AMS-02 experiment will give a stringent limit on the wino mass. With a few years of data taking, the limit will be increased, at least, up to 1 TeV~\cite{Evoli:2011id}. The LHC experiment will also provide an opportunity to directly search for the wino dark matter in near future. Using the pair production of the wino associated with a jet through electroweak interactions and utilizing the disappearing track of the charged wino~\cite{Direct wino production at LHC}, the limit on the wino mass is expected to be increased up to $500\,{\rm GeV}$, though the limit is currently very mild as $m_{\tilde w} > 100\,{\rm GeV}$, as reported in reference~\cite{ATLAS-CONF-2012-111}.

\section{Imprints on the Small-Scale Structure}
\label{sec:SmallScale}
Now, let us discuss possible imprints on the small-scale structure 
of the non-thermally produced wino dark matter.
For that purpose, we discuss the energy distribution
of the neutral wino in detail.
As we will see, a sizable fraction of the neutral wino keeps its energy
from the scattering process with the thermal background, which 
leads to a rather small free-streaming scale of the wino dark matter which can be observed
in the future $21{\rm cm}$ line survey.

\subsection{The wino energy distribution at the gravitino decay}
First, let us calculate the  energy distribution of the 
wino produced by the decays of the gravitino.
The gravitino decays into the wino through two-body decay modes.
In addition, the gravitino decays into the wino through the modes into heavier gauginos, 
i.e. the gluino and the bino.%
\footnote{
As we have mentioned above, we assume that all the decay
modes of the gravitino into the sfermions and the higgsinos are closed.
}
The wino has a line spectrum at the energy 
$E_{\tilde w} \simeq m_{3/2}/2$ from the two-body decay
and continuous spectra from the cascade decays which lead
to the energy distribution,
\begin{eqnarray}
\frac{1}{\Gamma_{3/2}} \frac{d\Gamma_{3/2}}{d E_{\tilde w}}
&=& 
\frac{8}{12}
\left\{
\frac{Br_{\tilde g \to \tilde w}}{\Gamma_{\tilde g\to \tilde w}}
\frac{d\Gamma_{\tilde g \to \tilde w}}{d E_{\tilde w}}
+
\int d{E_{\tilde b}}\,
 \frac{Br_{\tilde g \to \tilde b}}{\Gamma_{\tilde g\to \tilde b}}\frac{d\Gamma_{\tilde g\to \tilde b}}{d E_{\tilde b}}
 \frac{1 }{\Gamma_{\tilde b\to \tilde w}}\frac{d\Gamma_{\tilde b\to \tilde w}}{d E_{\tilde w}}
\right\}\cr
&&
+ 
\frac{1}{12}
\left\{
 \frac{1}{\Gamma_{\tilde b\to \tilde w}}\frac{d\Gamma_{\tilde b \to \tilde w}}{d E_{\tilde w}}
\right\}
+
\frac{3}{12} \delta(E_{\tilde w}-m_{3/2}/2)\ .
\end{eqnarray}
Here, $\Gamma$'s and $Br$'s denote the decay rates and the branching ratios 
of the gauginos which are given in the appendix\,\ref{sec:decay}.
The first and the third terms are the spectra from 
the cascade decays of $\tilde G \to  \tilde g \to \tilde w$
and $\tilde G \to  \tilde b \to \tilde w$ respectively, 
and the second term the spectrum from a cascade decay of $\tilde G \to  \tilde g \to \tilde b\to \tilde w$.

\begin{figure}[tb]
 \begin{minipage}{.49\linewidth}
  \begin{center}
  \includegraphics[width=\linewidth]{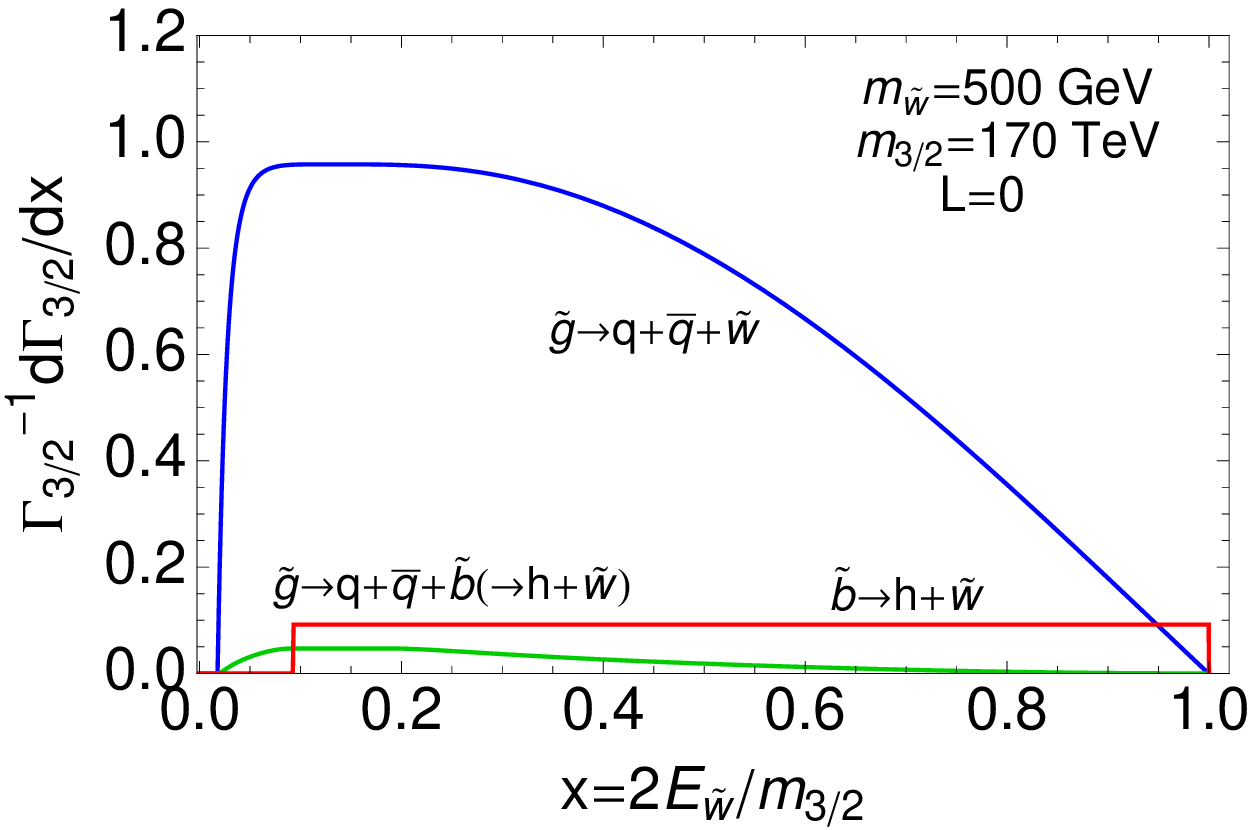}
  \end{center}
 \end{minipage}
 \begin{minipage}{.49\linewidth}
  \begin{center}
  \includegraphics[width=\linewidth]{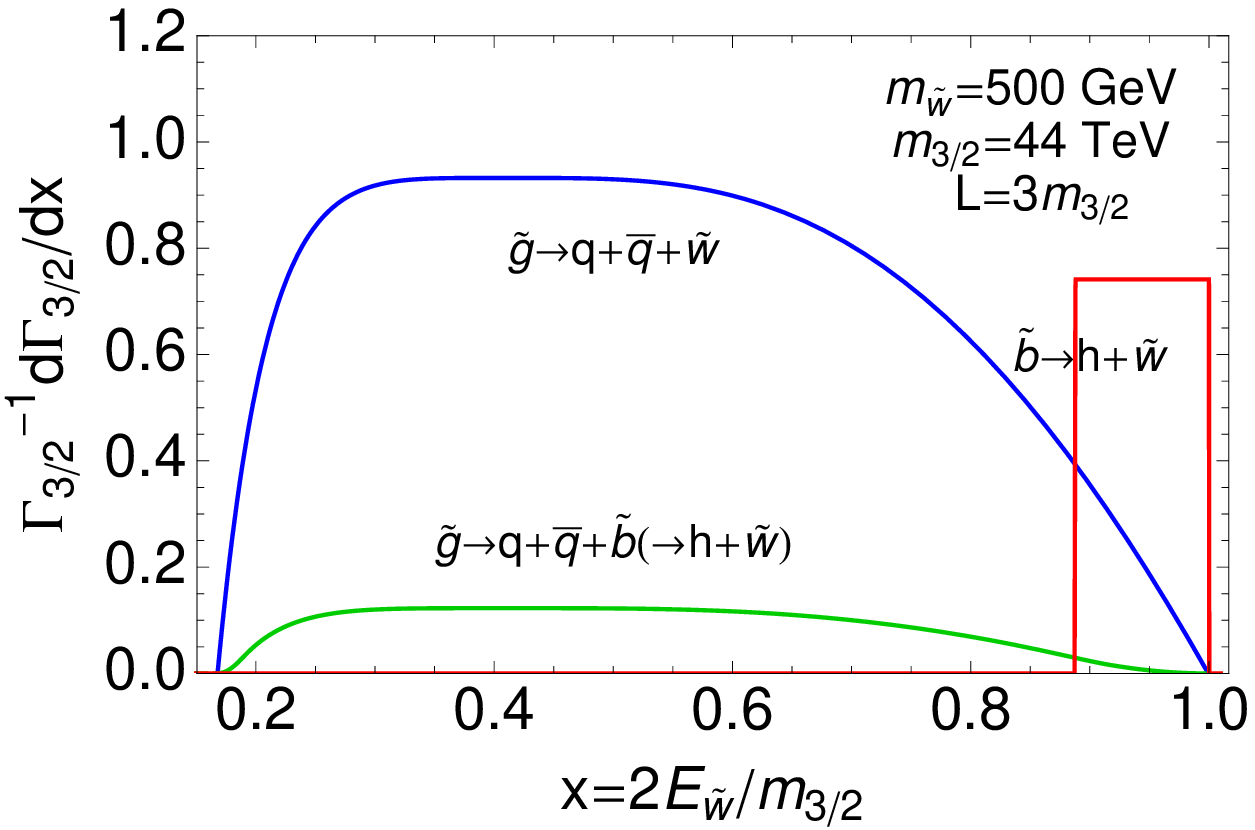}
  \end{center}
 \end{minipage}
\caption{\sl \small
The energy distribution of the wino produced by the cascade decays
of the gravitino
for given $m_{\tilde w}$ and  $L$.
We have taken $m_{\tilde w} = 500\,$GeV, $L=0$ (left panel) and
$L=3m_{3/2}$ (right panel), $\arg[m_{\tilde w}/m_{\tilde g}] = \pi$ and $\arg[m_{\tilde b}/m_{\tilde g}] = \pi$.
}
\label{fig:dGdx}
\end{figure}

In Fig.\,\ref{fig:dGdx}, we show the energy distribution of the wino produced by 
the cascades decay of the gravitino for given $m_{\tilde w}$ and $L$.
In the figure, we have taken $m_{\tilde w} = 500\,$GeV
and $L=0$ (left panel) and $L=3m_{3/2}$ (right panel). Here, we have taken $\arg[m_{\tilde w}/m_{\tilde g}] = \pi$ and $\arg[m_{\tilde b}/m_{\tilde g}] = \pi$
although the effects of the relative phases to our estimation
of possible imprints on the small-scale structure 
of the non-thermally produced wino dark matter are negligible (see the following discussions).
The figure shows that the peak of the wino energy distribution can be much 
smaller than the $m_{3/2}/2$ and has a low energy tail.

Notice that the peak position is higher for $L=3m_{3/2}$ for a given gravitino mass.
This is because the wino mass is closer to the gluino mass for $L=3 m_{3/2}$ 
than in the case of $L=0$, and hence, the wino carries away most of the gluino energy.
For a given wino mass, on the contrary, the energy of the wino is 
softer for $L=3m_{3/2}$ than the one for $L=0$, since
the gravitino mass is smaller for a larger $L$.

\subsection{Scattering processes with the thermal background}
\label{subsec:scatteringprocess}
The energetic winos produced by the decays of the gravitinos lose their energy through interactions with the thermal background, which consists of the electrons and positrons (e) and the neutrinos ($\nu_{l},\,l=e,\mu,\tau$) at the cosmic temperature of interest,  $T \sim 0.5 - 100 {\rm MeV} $.
In Refs.\,\cite{Hisano:2000dz, Arcadi:2011ev}, they study the interactions of the winos with the thermal background in detail. 
In this subsection, we summarize the relevant interactions for our purpose.

The charged winos mainly lose their energy via the Coulomb scattering with electrons and positrons in the thermal background. 
The energy loss rate at the cosmic temperature $T$ takes the form\,\cite{Reno:1987qw},
\begin{eqnarray}
-\frac{dE_{{\tilde w}^{\pm}}}{dt}=\frac{\pi \alpha^2 T^2}{3}\Lambda \left(1-\frac{m^2_{\tilde w}}{2E^2_{{\tilde w}^{\pm}}}\ln \left( \frac{E_{{\tilde w}^{\pm}}+p_{{\tilde w}^{\pm}}}{E_{{\tilde w}^{\pm}}-p_{{\tilde w}^{\pm}}} \right) \right)
\end{eqnarray}
with the fine-structure constant $\alpha$ and the Coulomb logarithm $\Lambda \sim O(1)$. 
Here, $E_{{\tilde w}^{\pm}}$ and $p_{{\tilde w}^{\pm}}$ are the physical energy and the physical momentum of the charged wino. 
The charged winos also turn into neutral winos by the decay of ${\tilde w}^{\pm} \to {\tilde w}^{0} + \pi^{\pm}$ and the inelastic scattering of ${\tilde w}^{\pm} + e \, (\nu_{e}) \to {\tilde w}^{0} + \nu_{e} \, (e)$. 
The decay is dominant\,\cite{Arcadi:2011ev} at the cosmic temperature of interest, and we neglect the inelastic scattering in the following. 
The average ratio of energy lost within one lifetime is given by, 
\begin{eqnarray}
\frac{\Delta E_{{\tilde w}^{\pm}}}{E_{{\tilde w}^{\pm}}}{\Big |}_{1-{\rm lifetime}} \simeq && 97 \Lambda \left(1 - \frac{m_\pi^2}{{\mit \Delta}m_{\tilde w}^2} \right)^{-1/2}
\left (\frac{T}{1 \, {\rm MeV}} \right)^2 \left(\frac{160 \, {\rm MeV}}{{\mit \Delta} m_{\tilde w}}\right)^3 \left (\frac{100 \, {\rm GeV}}{m_{\tilde w}} \right) \notag \\
&& \quad \times \left(1-\frac{m^2_{\tilde w}}{2E^2_{{\tilde w}^{\pm}}}\ln \left( \frac{E_{{\tilde w}^{\pm}}+p_{{\tilde w}^{\pm}}}{E_{{\tilde w}^{\pm}}-p_{{\tilde w}^{\pm}}} \right) \right)\ .
\end{eqnarray}
The charged winos lose most of their energy within one lifetime due to their long life time. 
Thus, the neutral winos produced via the decay of the charged winos are ``cold''.
Since the gravitino decays into the neutral and charged wino equally, at least two-thirds of dark matter particles are ``cold''.

\begin{figure}[tb]
 \begin{minipage}{.49\linewidth}
 \begin{center}
 \includegraphics[width=\linewidth]{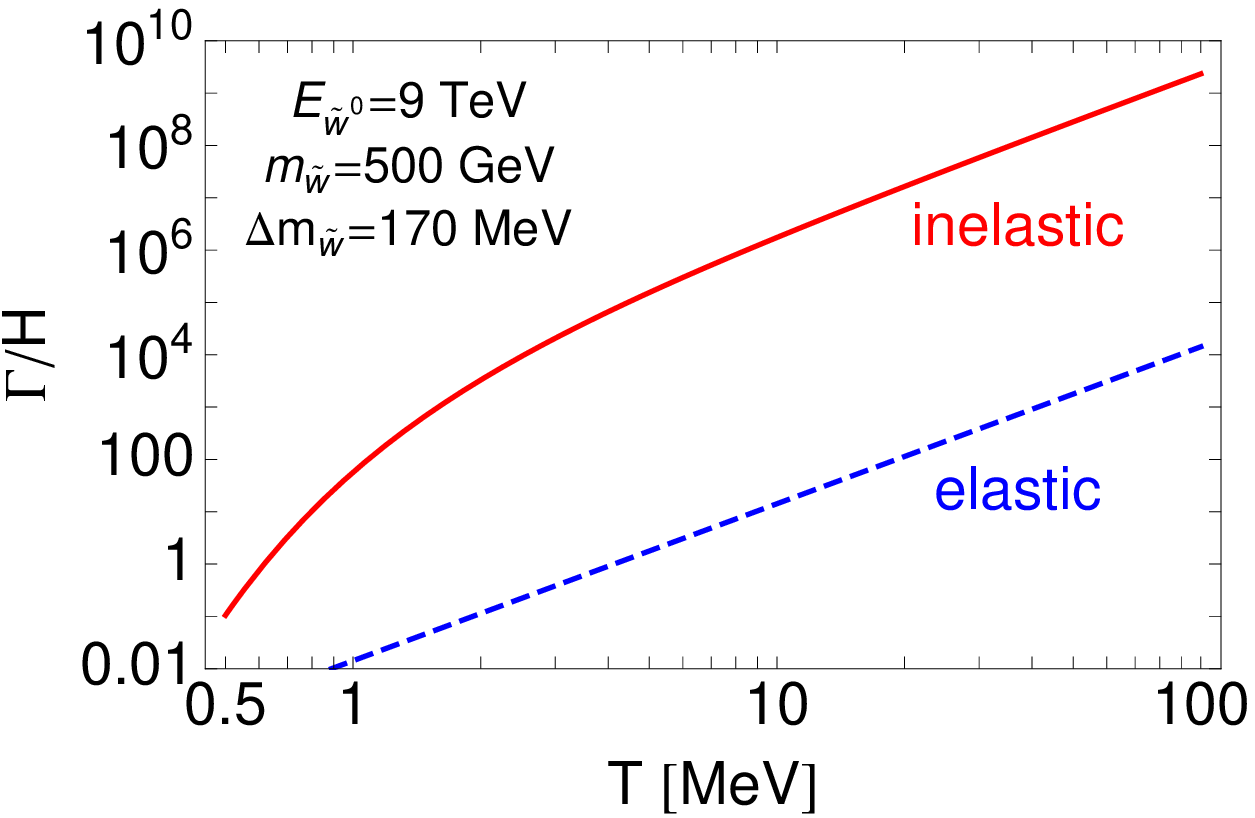}
  \end{center}
  \end{minipage}
 \begin{minipage}{.49\linewidth}
 \begin{center}
 \includegraphics[width=\linewidth]{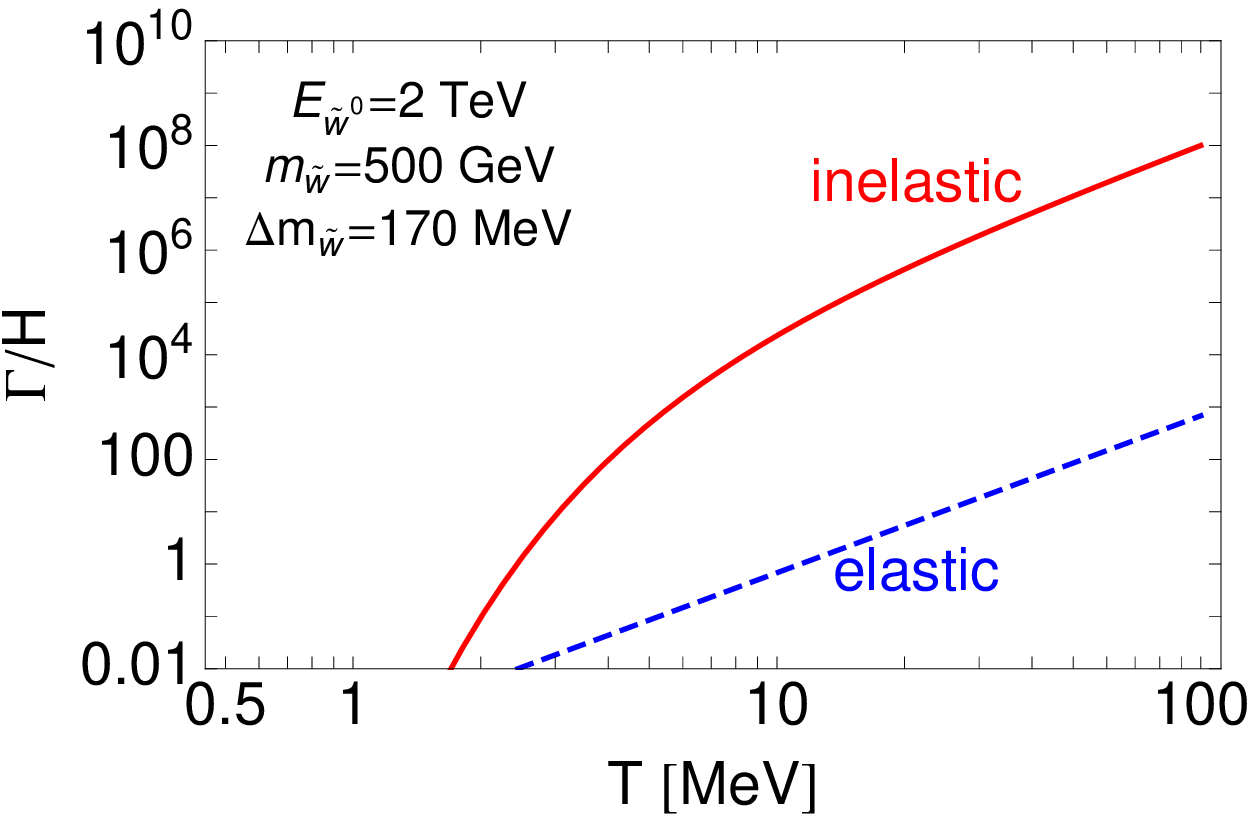}
  \end{center}
 \end{minipage}
\caption{\sl \small
The reaction rates of the neutral wino normalized by Hubble parameter as functions of the cosmic temperature. Here, we plot the reaction rates of the inelastic scattering (solid lines) and the elastic scattering (dashed lines), taking $m_{\tilde w} = 500 \, {\rm GeV}$ and ${\mit\Delta} m_{\tilde w}=170 \, {\rm MeV}$ for both  $E_{{\tilde w}^{0}} = 9 \, {\rm TeV}$ (left panel) and $E_{{\tilde w}^{0}} = 2 \, {\rm TeV}$ (right panel). }
\label{fig:winoreaction}
\end{figure}

The neutral wino doesn't have any elastic energy loss process at the tree level since we assume the $\mu-$parameter, the $B-$parameter and the sfermion masses are of the order of the gravitino mass.
Thus, the neutral winos can lose their energy via the elastic scattering at loop level and the inelastic scattering of ${\tilde w}^{0} + e \, (\nu_{e}) \to {\tilde w}^{\pm} + \nu_{e} \, (e)$.
We consider the elastic scattering at the one-loop level and study it in detail in appendix\,\ref{loopscatteing}. 
The rates of these processes are given by,
\begin{eqnarray}
&&\Gamma_{{\tilde w}^{0}, \, {\rm elastic}}= \frac{135}{\pi^3} \zeta(5) g_{\rm loop}^2 \! \left(\frac{m_{W}^2}{m_{\tilde w}^2} \right) G_{F}^4 T^5 m_{W}^4 \frac{E_{{\tilde w}^{0}}^2}{m_{\tilde w}^2} \left( 1+\frac{p_{{\tilde w}^{0}}^2}{E_{{\tilde w}^{0}}^2} \right)\ ,  \\
&&\Gamma_{{\tilde w}^{0}, \, {\rm inelastic}}=\frac{8}{\pi^3} G_{F}^2 T^5 \frac{(E_{{\tilde w}^{0}}+p_{{\tilde w}^{0}})^4}{m_{{\tilde w}}^2E_{{\tilde w}^{0}}p_{{\tilde w}^{0}}} \left(6+2\frac{m_{\tilde w}}{E_{{\tilde w}^{0}}+p_{{\tilde w}^{0}}}\frac{{\mit \Delta} m_{\tilde w}}{T} \right) \notag \\
&& \qquad \qquad  \qquad \times\exp \left(-\frac{m_{\tilde w}}{E_{{\tilde w}^{0}}+p_{{\tilde w}^{0}}}\frac{{\mit \Delta} m_{\tilde w}}{T} \right)
\end{eqnarray}
with the Riemann zeta function $\zeta(x)$, the Fermi constant $G_{F}$ and the mass of the weak boson $m_{W}$.
Here, $E_{{\tilde w}^{0}}$ and $p_{{\tilde w}^{0}}$ are the physical energy and the physical momentum of the neutral wino. 
The function $g_{\rm {loop}}(x)$ is given in appendix\,\ref{loopscatteing}. 
Notice that the last factor of the inelastic scattering rate in the above expression represents the Boltzmann suppression.
In Fig.\,\ref{fig:winoreaction}, we plot the reaction rates normalized by the Hubble parameter. The figure shows that the inelastic scattering become inefficient quickly at low temperature by the Boltzmann suppression, but in the relevant region of the cosmic temperature in which $\Gamma / H \gg 1$, the inelastic scattering dominates the elastic scattering. The inelastic scattering rate is higher for more energetic neutral wino since the inelastic scattering originates from the higher dimensional operator, and moreover, more energetic neutral wino can overcome the mass splitting between the charged and the neutral wino more easily.

Once the neutral winos turn into charged winos by the inelastic scattering, the charged winos lose most of their energy by the Coulomb scattering with the thermal background and turn back into neutral winos as discussed above.
Therefore, the energetic neutral winos produced by the decays of the gravitinos become ``warm'' only if they are directly produced by the decays of the gravitinos and they don't undergo the inelastic scattering since then. We call such neutral wino the ``warm'' neutral wino and other neutral wino the ``cold'' neutral wino in the following.

\subsection{The present momentum distribution of the ``warm'' neutral wino}
Now we describe the Boltzmann equation which determines the time evolution of the momentum distribution of the ``warm'' neutral wino $f_{\rm w}(p_{\rm w}, t)$ where $p_{\rm w}$ and $t$ denote the physical momentum of the ``warm'' neutral wino and the cosmic time.
Throughout this paper, we normalize the momentum distribution of the ``warm'' neutral wino by the present number density of the whole wino dark matter.
Following the discussion of the previous subsections, the Boltzmann equation of the ``warm'' neutral wino does not depend on the charged wino or the ``cold'' neutral wino and takes the form,
\begin{eqnarray}
&&\frac{\partial}{\partial t}f_{\rm w}(p_{\rm w}, t) - H p_{\rm w} \frac{\partial}{\partial p_{\rm w}}f_{\rm w}(p_{\rm w}, t) \notag \\
&& \quad = \frac{1}{3}\, \frac{d\Gamma_{3/2}}{d^3 p_{\rm w}}\frac{a(t_{0})^3}{a(t)^3}e^{-\Gamma_{3/2} t} - \Gamma_{{\tilde w}^{0}, \, {\rm inelastic}} \, f_{\rm w}(p_{\rm w}, t)
\end{eqnarray}
where $a(t)$ is the scale factor and $t_{0}$ is the present time. Here, $d\Gamma_{3/2} / d^3 p_{\rm w}$ relates to $d\Gamma_{3/2} / d E_{\rm w}$ as
\begin{eqnarray}
\frac{4 \pi p_{\rm w}^2}{(2\pi)^3} \, \frac{E_{\rm w}}{p_{\rm w}} \, \frac{d\Gamma_{3/2}}{d^3 p_{\rm w}} = \frac{d\Gamma_{3/2}}{d E_{\rm w}}
\end{eqnarray}
where $E_{\rm w}$ denote the physical energy of the ``warm'' neutral wino. In the Boltzmann equation of the ``warm'' neutral wino, we have ignored the thermal motion of the gravitinos since the gravitinos are nonrelativistic at the decay time. 

\begin{figure}[tb]
 \begin{minipage}{.49\linewidth}
 \begin{center}
 \includegraphics[width=\linewidth]{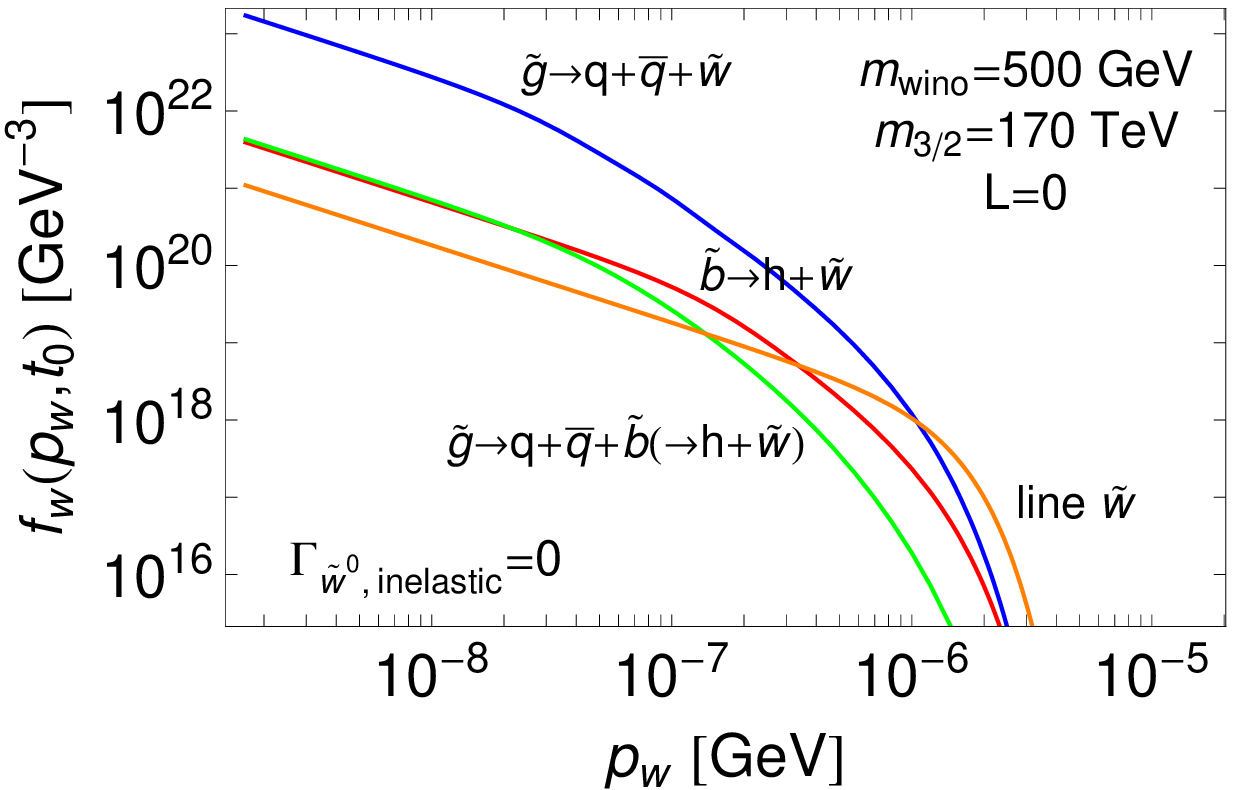}
  \end{center}
  \end{minipage}
 \begin{minipage}{.49\linewidth}
 \begin{center}
 \includegraphics[width=\linewidth]{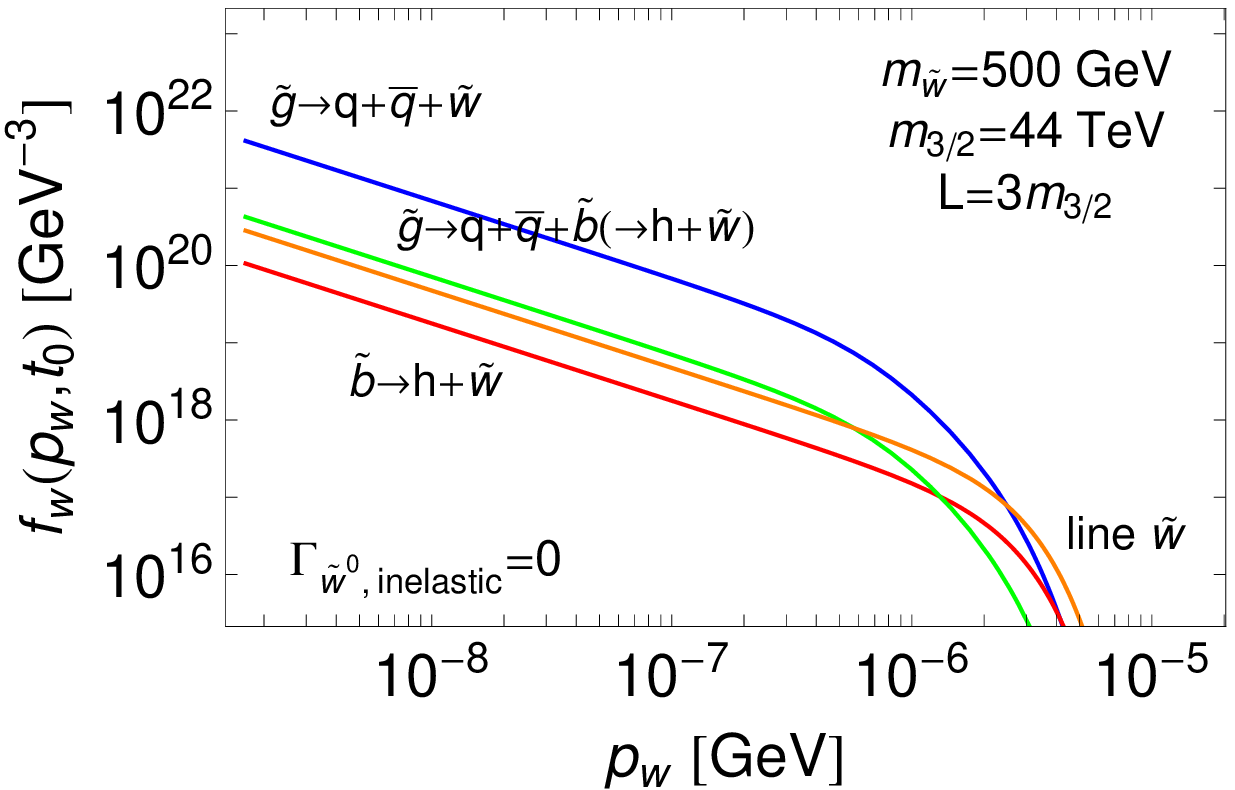}
  \end{center}
 \end{minipage}
  \begin{minipage}{.49\linewidth}
 \begin{center}
 \includegraphics[width=\linewidth]{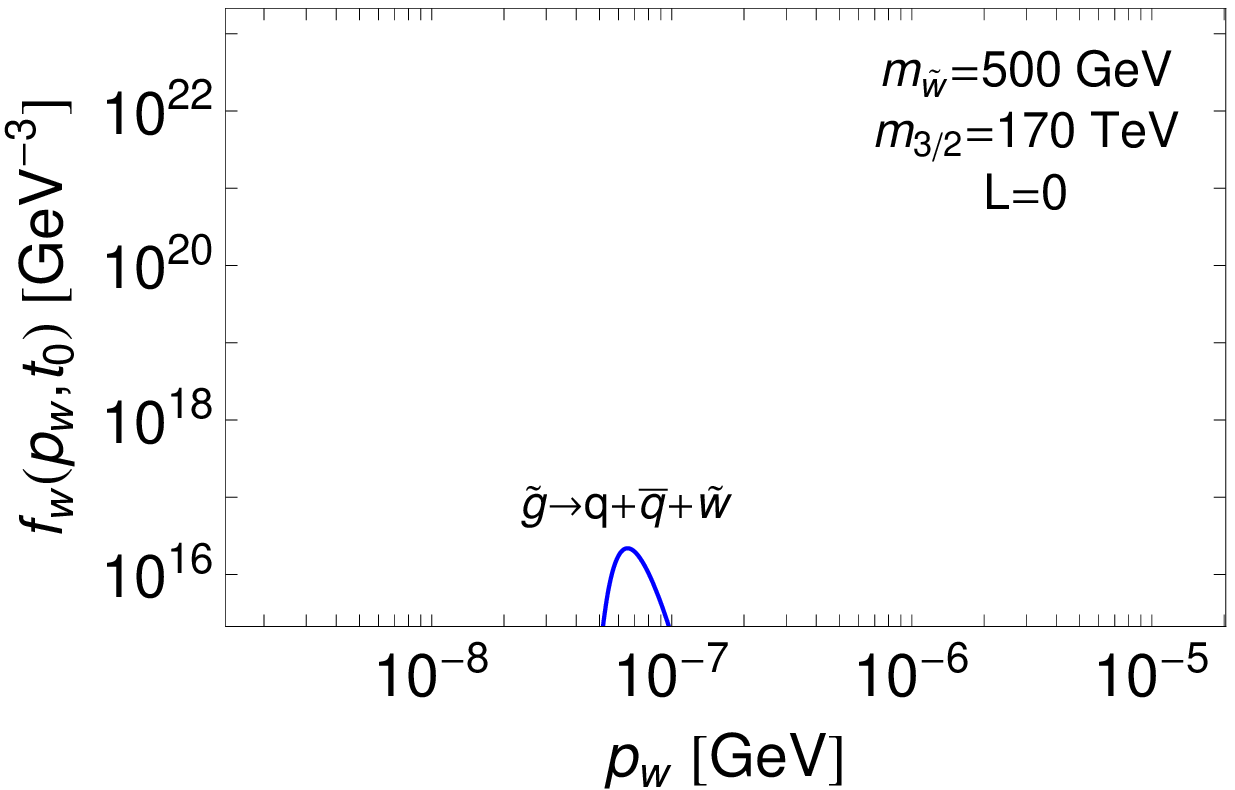}
  \end{center}
  \end{minipage}
 \begin{minipage}{.49\linewidth}
 \begin{center}
 \includegraphics[width=\linewidth]{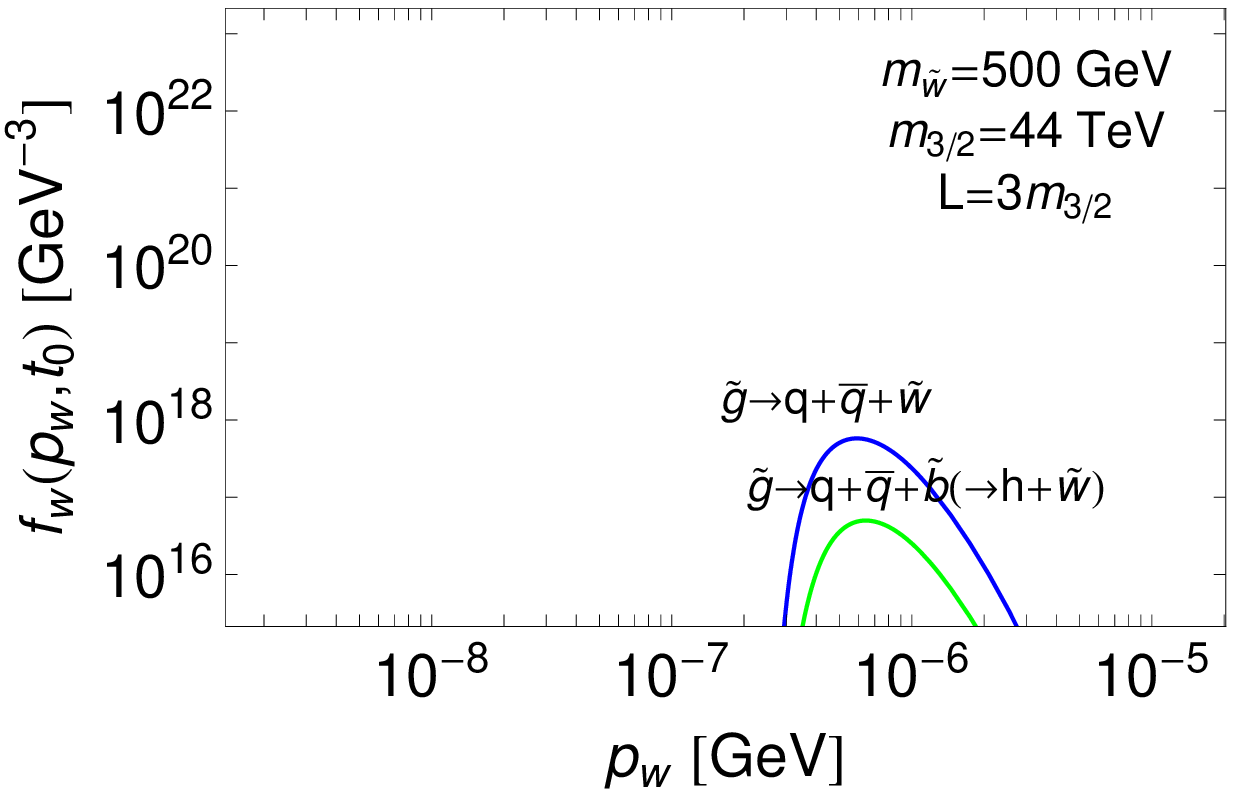}
  \end{center}
 \end{minipage}
\caption{\sl \small
The present momentum distribution of the ``warm'' neutral wino (lower panels). We also plot the present momentum distribution obtained by solving the Boltzmann equation without the inelastic scattering term for comparison (upper panels). Here, we have taken $m_{\tilde w} = 500\,$GeV, $L=0$ (left panels) and $L=3m_{3/2}$ (right panels) as in Fig.\,\ref{fig:dGdx}.}
\label{fig:winospectrum}
\end{figure}

By solving the Boltzmann equation numerically, we obtain the present momentum distribution of the ``warm'' neutral wino $f_{\rm w}(p_{\rm w}, t_{0})$. It should be noted that at the present time, $p_{\rm w}$ equals the comoving momentum of the ``warm'' neutral wino. In Fig.\,\ref{fig:winospectrum}, we plot the present momentum distribution of the ``warm'' neutral wino for the same parameters as in Fig.\,\ref{fig:dGdx} (lower panels). For comparison, in the upper panels we also plot the present momentum distribution of the ``warm'' neutral wino obtained by solving the Boltzmann equation without the inelastic scattering term, that is, letting $\Gamma_{{\tilde w}^{0}, \, {\rm inelastic}}=0$. When we ignore the inelastic scattering of the neutral wino, we can estimate the typical comoving momentum of the "warm" neutral wino as,
\begin{eqnarray}
p_{{\rm w},\,{\rm typical}} (t_{0}) = p_{\rm cm}\, a(t_{d})/a(t_{0})
\end{eqnarray}
where $p_{\rm cm}$ is the center-of-mass momentum and $t_{d}$ is the time of the gravitino decay, at which $T=T_{d}$.
In the present scenario, noting $p_{\rm cm}\simeq m_{3/2}/2$ and $a(t_{d})/a(t_{0}) \simeq (4/11)^{1/3}\,T_{0}/T_{d}$ where $T_{0}$ is the present temperature of the cosmic microwave background, the typical present momentum of the ``warm'' neutral wino is given by,
\begin{eqnarray}
p_{{\rm w},\,{\rm typical}} (t_{0}) \simeq 2.1 \times 10^{-6}\,{\rm GeV}\times \left( \frac{m_{3/2}}{100\,{\rm TeV}} \right)^{-1/2}\ .
\label{eq:typicalmomentum}
\end{eqnarray}
The typical present momentum of the ``warm'' neutral wino is larger for smaller gravitino mass due to later decays of lighter gravitino and we can see this relation in the upper panels of Fig.\,\ref{fig:winospectrum}.
When we consider the inelastic scattering of the neutral wino, the inelastic scattering turns a part of the neutral winos produced by the decays of the gravitinos into the ``cold'' neutral winos. In particular, almost all of  the neutral winos that are produced from the cascade decay of  $\tilde G \to  \tilde b \to \tilde w$ and the direct decay of $\tilde G \to  \tilde w$ turn into the ``cold'' neutral winos, because the winos produced by these decay modes are more energetic than the winos produced by the cascade decays of  $\tilde G \to  \tilde g \to \tilde w$ and $\tilde G \to  \tilde g \to \tilde b \to \tilde w$ as we can see in Fig.\,\ref{fig:dGdx}. 
The abundance of softer ``warm'' neutral winos as well as harder ``warm'' neutral winos is reduced when we consider the inelastic scattering.
This is because softer ``warm'' neutral winos are produced at the higher redshift, when the inelastic scattering rate is higher.

\subsection{The ratio of the ``warm'' neutral wino and the free-streaming scale of the wino dark matter}
According to the current understanding of the cosmology, the structure formation of the Universe occurs as a result of the mutual evolution of the primordial fluctuations of the gravitational potential and the energy density of dark matter. Since the ``warm'' neutral winos have high-velocity and can clime over the gravitational potential, they do not contribute to the structure formation of the Universe and reduce the power spectrum of the matter density fluctuations in comparison with the $\Lambda$CDM model.

When we discuss the imprints on the structure formation of the wino dark matter, we consider two indices. One is the ratio of the ``warm'' neutral wino to the whole wino dark matter, $r_{\rm warm}$. Smaller value of $r_{\rm warm}$ indicates smaller imprints on the matter power spectrum.  Another is the free-streaming scale which is defined as the comoving Jeans scale at the matter-radiation equality time and it takes the form, 
\begin{eqnarray}
k_{\rm fs}= a \sqrt{\frac{4 \pi G \rho_{\rm mat}}{\langle v^2 \rangle}} {\bigg |}_{a=a_{\rm eq}}
\end{eqnarray}
where $G=(8\pi M_{\rm Pl}^2)^{-1}$, $\rho_{\rm mat}$ and $\langle v^2 \rangle$ are the gravitational constant, the matter energy density of the Universe and the mean square of the velocity of dark matter, respectively. The primordial fluctuations with the wavelength above the free-streaming scale hold the imprints by the ``warm'' natural wino.
In the conventional warm dark matter scenario suggested as a solution of the so-called ``small-scale crisis''\,\cite{warmdarkmatter}, the ratio of the warm dark matter to the whole dark matter of the Universe is assumed to be one and the free-streaming scale is assumed to be around $100\,\rm{Mpc^{-1}}$. In the present wino dark matter scenario, however, $r_{\rm warm}$ does not exceed one-third, and hence, the wino dark matter is not the conventional warm dark matter. 

We calculate these two indices using the momentum distributions obtained in the previous subsection (see Fig.\,\ref{fig:winospectrum}) and summarize the results in Table.\ref{table:warmparameters}. In this table, we also show the results obtained when we ignore the inelastic scattering of the neutral wino, $\Gamma_{{\tilde w}^{0}, \, {\rm inelastic}}=0$, in the parentheses. We can confirm the ratio of the ``warm'' neutral winos to the whole wino dark matter is one-third when we ignore the inelastic scattering as mentioned above. 
For $L=3m_{3/2}$ the value of $r_{\rm warm}$ is smaller and the value of $k_{\rm fs}$ is larger to indicate the wino dark matter is ``colder'' than in the case of $L=0$.
This is because for heavier gravitino the value of  $p_{{\rm w},\,{\rm typical}} (t_{0})$ is smaller and the value of $\Gamma_{{\tilde w}^{0}, \, {\rm inelastic}}$ is larger.

\begin{table}[tb]
 \caption{\sl \small
 The ratio of the ``warm'' neutral wino to the whole wino dark matter $r_{\rm warm}$ and the free-streaming scale of the wino dark matter $k_{\rm fs}$, for $m_{\tilde w} = 500\,$GeV and $L=0$ and $L=3m_{3/2}$ as in Fig.\,\ref{fig:dGdx}. We also show the results for $\Gamma_{{\tilde w}^{0}, \, {\rm inelastic}}=0$ in the parentheses.
  }
 \label{table:warmparameters}
  \begin{center}
  \begin{tabular}{|c||c|c|} \hline
  parameters & $r_{\rm warm}$ & $k_{\rm fs}\,[\rm Mpc^{-1}]$ \\ \hline \hline
  $m_{\tilde w}=500\,{\rm GeV},\,m_{3/2}=170\,{\rm TeV},\,L=0$  & $1.5\times10^{-7}\,(0.33)$ & $4.5 \times10^{7}\,(2.7 \times 10^{3})$ \\ \hline
  $m_{\tilde w}=500\,{\rm GeV},\,m_{3/2}=44\,{\rm TeV},\,L=3 m_{3/2}$  & $0.016\,(0.33)$ & $7.5 \times10^{3}\,(1.2 \times 10^{3})$ \\ \hline
  \end{tabular}
 \end{center}
\end{table}

\begin{figure}[tb]
 \begin{minipage}{.49\linewidth}
 \begin{center}
 \includegraphics[width=\linewidth]{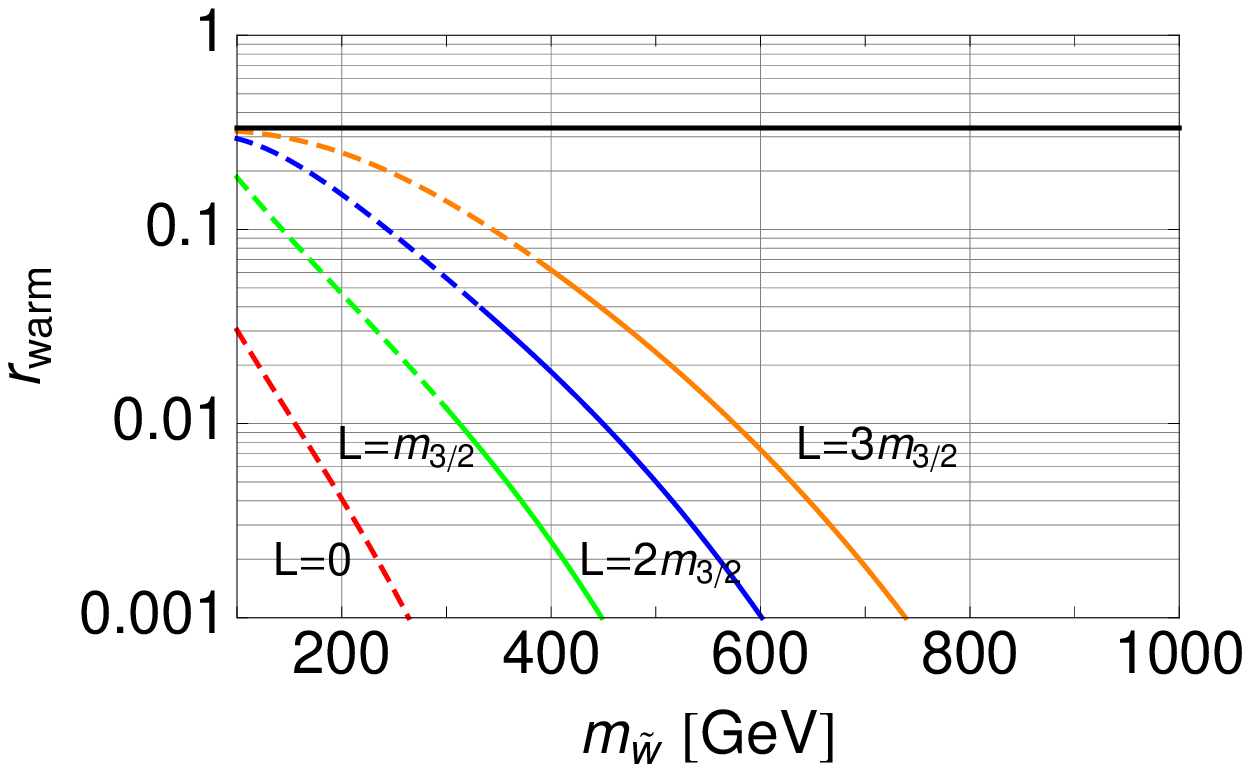}
  \end{center}
  \end{minipage}
 \begin{minipage}{.49\linewidth}
 \begin{center}
 \includegraphics[width=\linewidth]{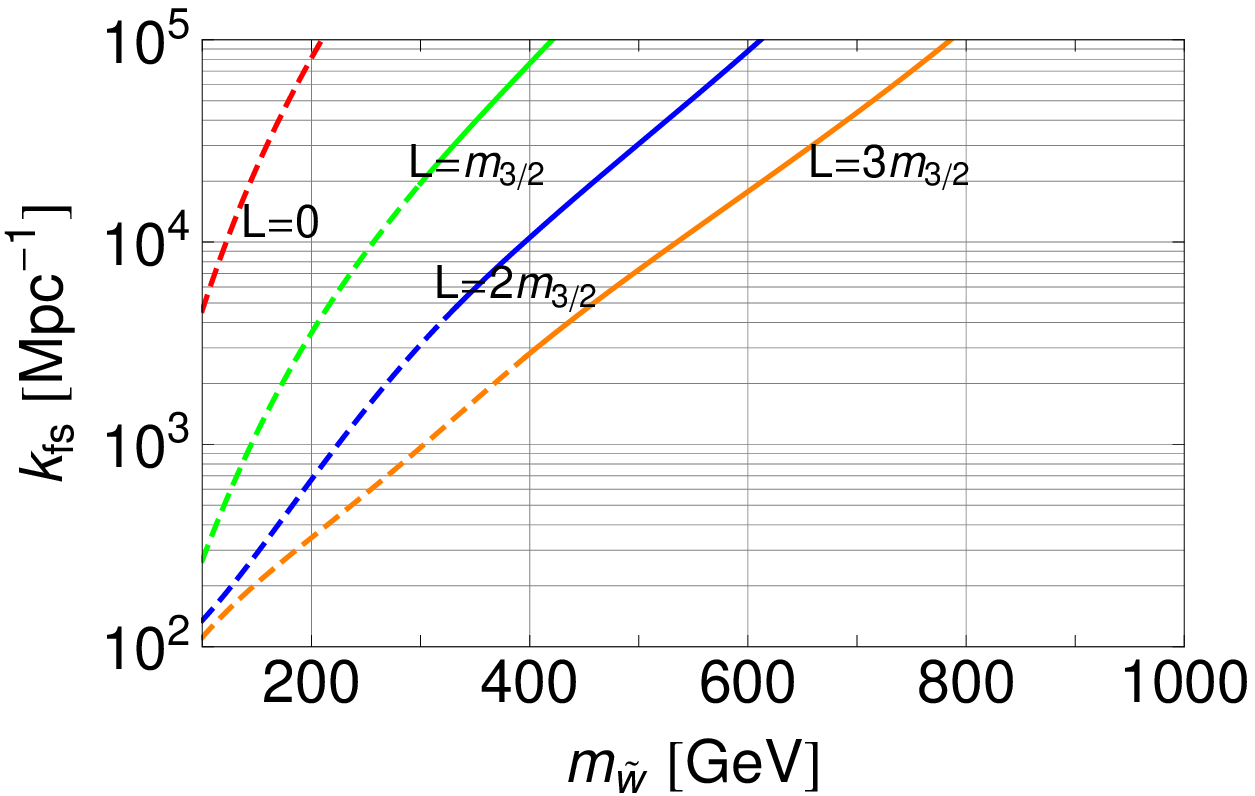}
  \end{center}
 \end{minipage}
\caption{\sl \small
The ratio of the ``warm'' neutral wino to the whole wino dark matter $r_{\rm warm}$ (left panel) and the free-streaming scale of the wino dark matter $k_{\rm fs}$ (right panel) as functions of the wino mass $m_{\tilde w}$ for $L=0,\,m_{3/2},\,2m_{3/2},\,3m_{3/2}$. The regions shown by the dashed lines are not favored by the collider experiments and the indirect detection searches.}
\label{fig:modelresults}
\end{figure}

In order to discuss the imprints on the matter power spectrum in the present scenario, we plot the values of $r_{\rm warm}$ (left panel) and $k_{\rm fs}$ (right panel) as functions of the wino mass $m_{\tilde w}$ for $L=0$,\,$m_{3/2}$,\,$2m_{3/2}$,\,$3m_{3/2}$ in Fig.\,\ref{fig:modelresults}. Since larger value of $L$ means smaller gravitino mass for a given wino mass, the wino dark matter for larger value of $L$ is ``warmer'' as we can see in the figure. 

\begin{figure}[tb]
 \begin{minipage}{.49\linewidth}
 \begin{center}
 \includegraphics[width=\linewidth]{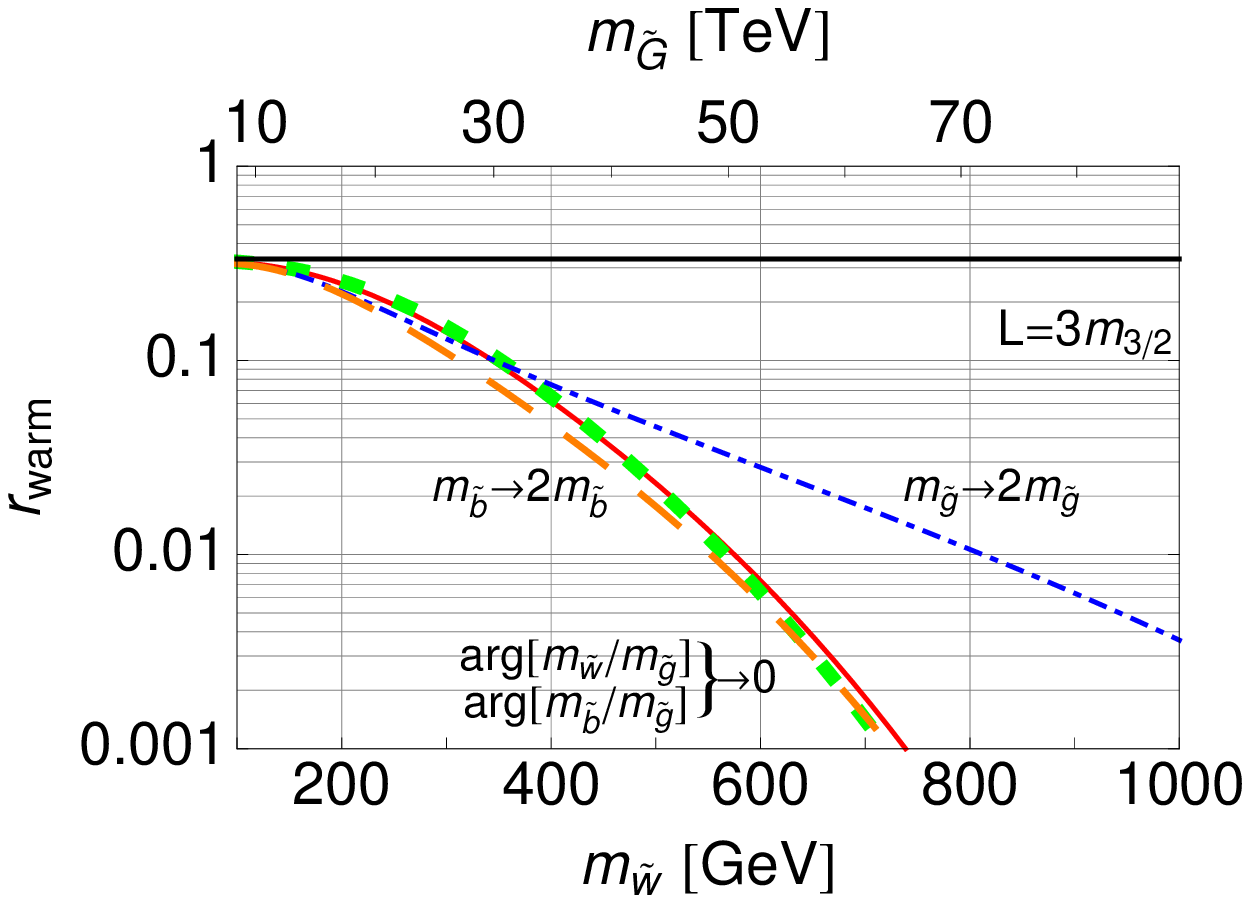}
  \end{center}
  \end{minipage}
 \begin{minipage}{.49\linewidth}
 \begin{center}
 \includegraphics[width=\linewidth]{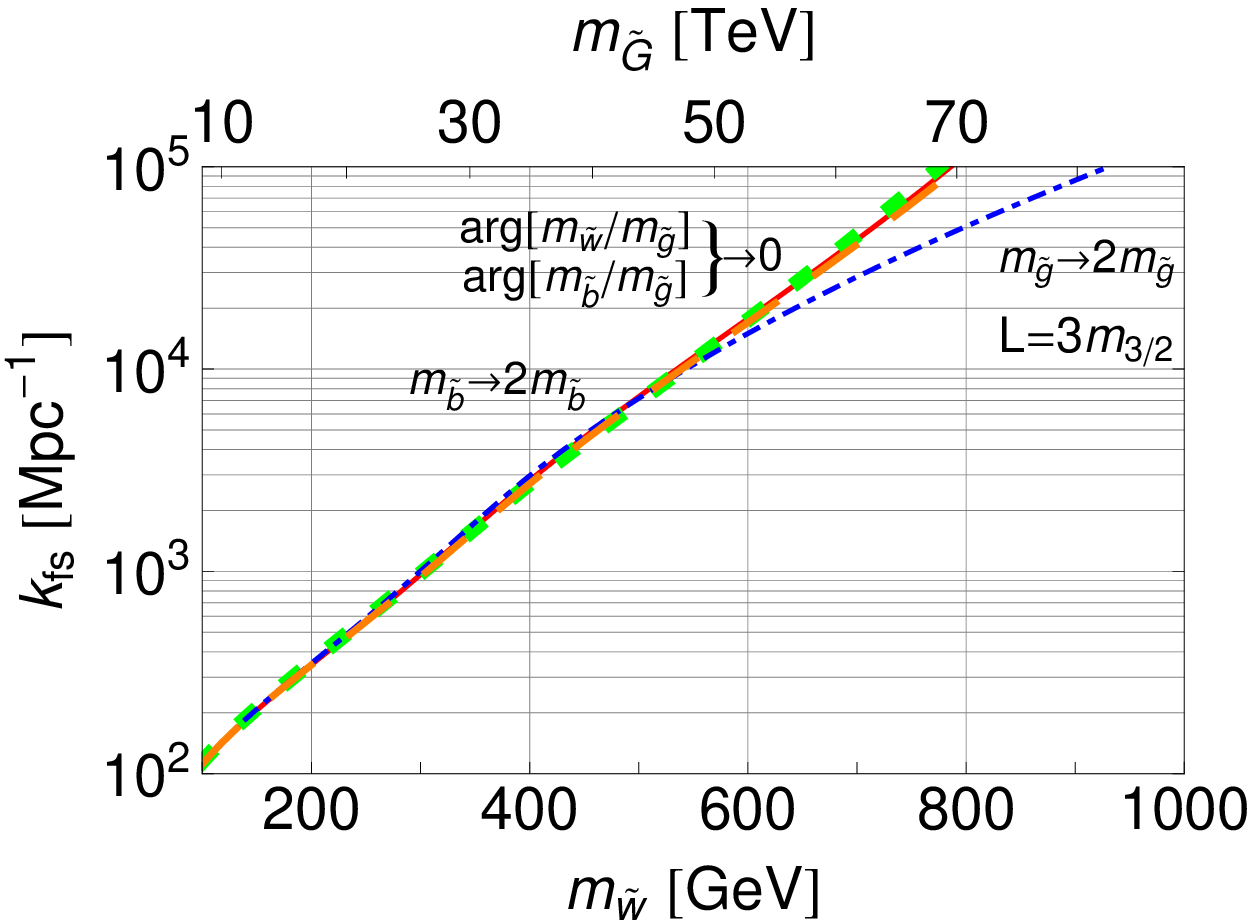}
  \end{center}
 \end{minipage}
\caption{\sl \small
The changes in the values of $r_{\rm warm}$ (left panel) and $k_{\rm fs}$ (right panel) for various values of $\arg[m_{\tilde w}/m_{\tilde g}]$, $\arg[m_{\tilde b}/m_{\tilde g}]$ (dotted lines), $m_{\tilde b}$ (dashed lines) and $m_{\tilde g}$ (dot dashed lines). Here, we have taken $L=3m_{3/2}$.}
\label{fig:changesintheindices}
\end{figure}

In the above discussion of this section, we have taken the relative phases between the gaugino mass parameters, $\arg[m_{\tilde w}/m_{\tilde g}]=\pi$ and $\arg[m_{\tilde b}/m_{\tilde g}]=\pi$, and we have also fixed the relations between the gaugino masses determined by given $m_{\tilde w}$ and $L$ in the present scenario.
Now, we vary the values of $\arg[m_{\tilde w}/m_{\tilde g}]$, $\arg[m_{\tilde b}/m_{\tilde g}]$, $m_{\tilde b}$ and $m_{\tilde g}$ to see their effects on the values of $r_{\rm warm}$ and $k_{\rm fs}$. 
In Fig.\,\ref{fig:changesintheindices}, we plot the values of $r_{\rm warm}$ (left panel) and $k_{\rm fs}$ (right panel) for various values of $\arg[m_{\tilde w}/m_{\tilde g}]$, $\arg[m_{\tilde b}/m_{\tilde g}]$ (dotted lines), $m_{\tilde b}$ (dashed lines) and $m_{\tilde g}$ (dot dashed lines). In the figure, we have taken $L=3m_{3/2}$. 
The figure shows the effects of the relative phases between the gaugino mass parameters, $\arg[m_{\tilde w}/m_{\tilde g}]$ and $\arg[m_{\tilde b}/m_{\tilde g}]$ are at most $10\,\%$ and negligible. We can also see the effects of the bino mass $m_{\tilde b}$ are small. 
On the other hand, larger gluino mass $m_{\tilde g}$ significantly enhance the value of $r_{\rm warm}$ and reduce the value of $k_{\rm fs}$, that is, make the wino dark matter ``warmer''. 
This is because the wino carries away smaller fraction of the gluino energy for larger mass difference between the gluino and the wino.
Therfore, in other models where the gluino mass for a given wino mass is lager than in the present model, the imprints on the matter power spectra are enhanced. 

As we have seen above, the wino dark matter is the so-called mixed dark matter\,\cite{Anderhalden:2012qt} such as massive neutrinos\,\cite{Lesgourgues:2006nd} rather than the warm dark matter.
The present constraints on the mixed and the warm dark matter come from the observations of the large-scale structure, especially Lyman-alpha cloud\,\cite{Boyarsky:2008xj}, while they can put the constraint on the mixed and warm dark matter with rather smaller free-streaming scale $k_{\rm fs}\ll 100\,{\rm Mpc^{-1}}$ than the typical free-streaming scale $k_{\rm fs} > 100\,{\rm Mpc^{-1}}$ in the present scenario.
We suggest that the future observations of the redshifted $21\,{\rm cm}$ line should be the most promising. In fact, the detectability of the neutrino mass in the future $21\,{\rm cm}$ line survey are discussed\,\cite{21cmneutrino}. The spatial fluctuations of the $21\,{\rm cm}$ line\ absorbed between $30\lesssim z \lesssim200$ can directly probe the linear matter density fluctuation with a comoving wavenumber larger than $100\,{\rm Mpc^{-1}}$ (wavelength smaller than $100\,{\rm kpc}$)\,\cite{Loeb:2003ya}. The detectability of the imprints of the non-thermal wino dark matter in the future $21\,{\rm cm}$ line survey should be studied, while it's beyond the scope of this paper.

\section{Conclusions}
The wino dark matter is highly motivated in scenarios with heavy scalars such as the pure gravity mediation model after
the discovery of a Higgs-like particle with a mass around $125\,{\rm GeV}$ at the LHC.
In this paper, we have studied how ``warm'' the wino dark matter is when it is non-thermally produced by the late time decays of the gravitinos.

The winos produced by the cascade decays of the gravitino have the energy of $E_{\tilde w}\sim 0.1\,m_{3/2}$.
The charged winos lose almost all of their evergy within one-lifetime through the electromagnetic interaction with the thermal background.
As a result, at least two-thirds of the wino dark matter are ``cold''. 
The neutral winos produced by the decays of the gravitino can turn into charged winos by the inelastic scattering.
The neutral winos which don't undergo the inelastic scattering become ``warm''.

As we have shown, a sizable fraction of the wino dark matter can be ``warm'' for the wino mass $m_{\tilde w}\simeq100-500\,{\rm GeV}$.
The imprints on the matter power spectra may provide further insight on the origin of dark matter via the future $21\,{\rm cm}$ line survey
in combination with the LHC experiments and the indirect dark matter searches.
Our calculations can be applied to generic wino LSP scenarios with the heavy sfermions and higgsinos.
It should be noted that for heavier gluino scenarios, the imprints on the matter power spectra are enhanced, while searches in the LHC experiments become difficult.
The detectability of the imprints of the non-thermal wino dark matter in the future $21\,{\rm cm}$ line survey is worthy of the future study.

Finally, we comment on a further application of this work.
The higgsino dark matter scenarios are suggested in the supersymmetric models\,\cite{Drees:1996pk, Fujii:2002kr}.
The neutral and the charged higgsinos produced by the decays of the gravitino can leave the imprints on the matter power spectra as well as the neutral and the charged winos. 
In the higgsino dark matter scenarios, the mass splittings between the charged and the neutral higgsinos are in a {\rm GeV} range, and hence, the lifetime of the charged higgsino is much shorter than the lifetime of the charged wino to make the higgsino dark matter ``warmer''.
On the other hand, the higgsino-gaugino mixing provides the elastic scattering of the higgsino at the tree level as well as the one-loop level to make the higgsino dark matter ``colder''.
A comprehensive study of the non-thermal higgsino dark matter will be given elsewhere\,\cite{Ibe:inprep}.

\section*{Acknowledgments}
The authors thank T.T. Yanagida for useful discussion at the early stage of the project. A.K. also thanks T. Takesako for providing the note of his 
previous work. 
This work is supported by Grant-in-Aid for Scientific research from the Ministry of Education, Science, Sports, and Culture (MEXT), Japan, No. 24740151 (M.I.), No. 22244021 and No. 23740169 (S.M.), JSPS Research Fellowships for Young Scientists (A.K.) and also by World Premier International Research Center Initiative (WPI Initiative), MEXT, Japan.

\appendix
\section{Partial Decay Widths of the Gauginos}
\label{sec:decay}
In this appendix, we calculate the partial decay widths of the 
gluino and the bino produced by the decay of the gravitino used in section\,\ref{sec:SmallScale}.
\subsection{Partial decay widths of the gluino}
The gluino decays into the winos and the bino through 
the three-body decays, $\tilde g\to q + \bar{q}+ \tilde{w}$.
The spin averaged and color summed partial decay width of the neutral wino mode 
is given by,
\begin{eqnarray}
\frac{d\Gamma_{\tilde g \to \tilde w} }{d E_{\tilde w}}&=&
 \frac{2}{3\cdot (8\pi)^3} \frac{g_3^2 g_2^2}{m_{\rm squark}^4}
\frac{m_{\tilde g}^6}{p_{\tilde g} E_{\tilde g}} 
\Big(
2 r_{\tilde w}
(3c_{\tilde w}   -2 r_{\tilde w} + 3 c_{\tilde w} r_{\tilde w}^2
)
(\e_{\tilde w, {\rm CM}, {\rm upper} }
-\e_{\tilde w, {\rm CM}, {\rm lower} })
 \cr 
&&+ 3 (1 - 2c_{\tilde w}  r_{\tilde w} + r_{\tilde w}^2)
(\e_{\tilde w, {\rm CM}, {\rm upper} }^2
-\e_{\tilde w, {\rm CM}, {\rm lower} }^2)\cr
&&
-\frac{8} {3}
(\e_{\tilde w, {\rm CM}, {\rm upper}}^3
-\e_{\tilde w, {\rm CM}, {\rm lower} }^3) \Big)\ ,
\end{eqnarray}
where $g$'s are the gauge coupling constants, $p_{\tilde g}$ is the size
of the three-dimensional momentum of the gluino,
$r_{\tilde w}$ is the mass ratio $r_{\tilde w}= m_{\tilde w}/m_{\tilde g}$,
and $c_{\tilde w}$ is the relative phase $c_{\tilde w} = \cos(\arg[m_{\tilde w}/m_{\tilde g}])$.
In our analysis, we neglect the masses of the standard model fermion
for simplicity.

In the above expression, we have introduced $\epsilon$'s which are defined by
\begin{eqnarray}
 \label{eq:lower2}
 \e_{\tilde w, {\rm CM}, \, {\rm lower}} &=& \frac{E_{\tilde g} E_{\tilde w} - p_{\tilde g} p_{\tilde w}}{m_{\tilde g}^2}\ , \\
  \label{eq:upper2}
 \e_{\tilde w, {\rm CM},\, {\rm upper}} &=&  \frac{m_{\tilde g}^2 + m_{\tilde w}^2}{2m_{\tilde g}^2}\ ,
\end{eqnarray}
for a highly boosted gluino, i.e. $m_{\tilde g} > (m_{\tilde g}^2 + m_{\tilde w}^2)/2 E_{\tilde g}$.
For $m_{\tilde g} > (m_{\tilde g}^2 + m_{\tilde w}^2)/2 E_{\tilde g}$,
 the wino energy in the boosted gluino frame is in between 
\begin{eqnarray}
 E_{\tilde w}^- &=& \frac{E_{\tilde g} (m_{\tilde g}^2 + m_{\tilde w}^2) 
- p_{\tilde g} (m_{\tilde g}^2 - m_{\tilde w}^2) 
 }{2 m_{\tilde g}^2}\ , \\
  E_{\tilde w}^+ &=& \frac{E_{\tilde g} (m_{\tilde g}^2 + m_{\tilde w}^2) 
+ p_{\tilde g} (m_{\tilde g}^2 - m_{\tilde w}^2) 
 }{2 m_{\tilde g}^2} \ .
\end{eqnarray}

For a less boosted gluino, i.e. $m_{\tilde w} < (m_{\tilde g}^2 + m_{\tilde w}^2)/2 E_{\tilde g}$, on the other hand, $\e$'s are defined by,
\begin{eqnarray}
\label{eq:lower1}
 \e_{\tilde w, {\rm CM}, \, {\rm lower}} &=& \frac{E_{\tilde g} E_{\tilde w} - p_{\tilde g} p_{\tilde w}}{m_{\tilde g}^2}\ , \\
 \label{eq:upper1}
 \e_{\tilde w, {\rm CM},\, {\rm upper}} &=&
\left\{
\begin{array}{ll}
\displaystyle{\frac{E_{\tilde g} E_{\tilde w} + p_{\tilde g} p_{\tilde w}}{m_{\tilde g}^2}}\  
   &  (E_{\tilde w} < E^-_{\tilde w}) \ ,
   \\
\displaystyle{ \frac{m_{\tilde g}^2 + m_{\tilde w}^2}{2m_{\tilde g}^2}}\  
   &   (E_{\tilde w} > E^-_{\tilde w}) \ ,
\end{array}
\right.
\end{eqnarray}
and the wino energy is in between $m_{\tilde w}$ and $E^+_{\tilde w}$ in
the boosted gluino frame.
In our application, the gluino produced by the gravitino decay is highly boosted, 
and hence, we use $\e$'s in Eqs.\,(\ref{eq:lower2}) and (\ref{eq:upper2}).
By integrating the above partial width in between 
$E_{\tilde w}^\pm$, we obtain the total decay width into the neutral wino, 
\begin{eqnarray}
\G_{\tilde g\to \tilde w}&=& \frac{4g_2^2 g_3^2}{3(16\pi)^3}\frac{m_{\tilde g}^6}{E_{\tilde g} 
m_{\rm squark}^4}
{\Big (}
  (1-r_{\tilde w}^2 ) \left(
1-7 r_{\tilde w}^2 -7 r_{\tilde w}^4 +r_{\tilde w}^6
+  2 c_{\tilde w} (r_{\tilde w} +10  r_{\tilde w}^3+ r_{\tilde w}^5 )
     \right)
   \cr
&& +  24 r_{\tilde w}^3 ( c_{\tilde w}-r_{\tilde w}c_{\tilde w}-r_{\tilde w}+ c_{\tilde w} r_{\tilde w}^2 ) \ln
  r_{\tilde w} 
{\Big )}\ .
\end{eqnarray}

Similarly, we obtain the decay width of the gluino into the bino via the 
three body decays,
\begin{eqnarray}
\G_{\tilde g\to \tilde b}&=& \frac{44g_1^2 g_3^2}{45(16\pi)^3}\frac{m_{\tilde g}^6}{E_{\tilde g} 
m_{\rm squark}^4}
{\Big (}
  \left(1-r_{\tilde b}^2\right) \left(
1-7 r_{\tilde b}^2 -7 r_{\tilde b}^4 +r_{\tilde b}^6
+  2 c_{\tilde b} (r_{\tilde b} +10  r_{\tilde b}^3+ r_{\tilde b}^5)
     \right)
   \cr
&& +  24 r_{\tilde b}^3 \left(c_{\tilde b}-r_{\tilde b}c_{\tilde b}-r_{\tilde b}+ c_{\tilde b} r_{\tilde b}^2\right) \ln
  r_{\tilde b}
{\Big )}\ ,
\end{eqnarray}
where $r_{\tilde b} = m_{\tilde b}/m_{\tilde g}$. 
The partial width is also obtained,
\begin{eqnarray}
\frac{d\Gamma_{\tilde g \to \tilde b} }{d E_{\tilde b}}&=&
 \frac{22}{45(8\pi)^3} \frac{g_3^2 g_1^2}{m_{\rm squark}^4}
\frac{m_{\tilde g}^6}{p_{\tilde g} E_{\tilde g}} 
{\Big (}
2 r_{\tilde b}
(3c_{\tilde b}   -2 r_{\tilde b} + 3 c_{\tilde b} r_{\tilde b}^2
)
(\e_{\tilde b, {\rm CM}, {\rm upper} }
-\e_{\tilde b, {\rm CM}, {\rm lower} })
 \cr 
&&+ 3 (1 - 2c_{\tilde b}  r_{\tilde b} + r_{\tilde b}^2)
(\e_{\tilde b, {\rm CM}, {\rm upper} }^2
-\e_{\tilde b, {\rm CM}, {\rm lower} }^2)\cr
&&
-\frac{8} {3}
(\e_{\tilde b, {\rm CM}, {\rm upper}}^3
-\e_{\tilde b, {\rm CM}, {\rm lower} }^3) {\Big)}\ .
\end{eqnarray}
Here, $\epsilon$'s  are obtained  
by replacing $\tilde w$ with $\tilde b$
in Eqs.\,(\ref{eq:lower2}) and (\ref{eq:upper2}),
and $c_{\tilde b} =\cos(\arg[m_{\tilde b}/m_{\tilde g}])$.
By remembering that the wino is a triplet, the branching ratio of the bino mode 
is given by,
\begin{eqnarray}
Br_{\tilde g\to\tilde b}=\frac{\G_{\tilde g \to\tilde  b}}{
3\G_{\tilde g \to \tilde w}+\G_{\tilde g \to \tilde b}
}\ ,
\end{eqnarray}
which is less than about $10$\,\% in the parameter space we are interested in.

\subsection{Partial decay widths of the bino}
The main decay modes of the bino are the two-body decay into
the neutral wino, $\tilde b\to h+ \tilde{w}$ and the ones
into the charged wino, $\tilde b\to W^\pm+ \tilde{w}^\mp$.
To calculate the decay widths, let us define the mass eigenstates 
of the neutralino mass matrix,
\begin{eqnarray}
M_{\rm neut} =
\left(
\begin{array}{cccc}
M_1 &0  &  - c_\b  s_W m_Z & s_\b   s_W  m_Z    \\
0 &M_2  & c_\b  c_W  m_Z   & -s_\b   c_W  m_Z   \\
  - c_\b  s_W  m_Z  &  c_\b  c_W m_Z  &  0 & -\m  \\
s_\b  s_W  m_Z &   -s_\b  c_W  m_Z  &  -\m &  0 
\end{array}
\right)\ ,
\end{eqnarray}
and the chargino mass matrix,
\begin{eqnarray}
 M_{\rm ch}  =  
\left(
\begin{array}{cc}
 M_2   & \sqrt{2} c_\b  c_W  m_Z   \\
   \sqrt{2} s_\b  c_W  m_Z    &    \mu
\end{array}
\right)\ .
\end{eqnarray}
where $M_{1,2}$ denotes the gaugino masses, $\m$ is the $\mu$-term, 
 $s_\b = \sin\beta$, $c_\b = \cos\beta$, and the subscript $W$ denotes the Weinberg angle.

The mass eigenstates up to $O(m_Z^2/\mu^2)$ are given by,
\begin{eqnarray}
\label{eq:bino0}
\tilde b &=&-\frac{ s_{2\b}s_{2W}m_Z^2}{2(M_1-M_2)\m}\, \chi^0_1  + \chi^0_2 + O(m_Z^2/\m^2) \ , \\
\label{eq:wino0}
\tilde w^0 &=&\chi^0_1 + \frac{ s_{2\b}s_{2W}m_Z^2}{2(M_1-M_2)\m}\, \chi^0_2  +  O(m_Z^2/\m^2) \ , \\
\tilde H_d^0 &=&
-\frac{s_\b  c_W m_Z}{\mu}\,\chi^0_1 
+\frac{s_\b s_W m_Z}{\mu}\,\chi^0_2 
+\chi^0_3 
 +  O(m_Z^2/\m^2) \ , \\
 \tilde H_u^0 &=&
 \frac{c_\b  c_W m_Z}{\mu}\,\chi^0_1 
- \frac{c_\b s_W m_Z}{\mu}\,\chi^0_2 
+\chi^0_4 
 +  O(m_Z^2/\m^2) \ , 
\end{eqnarray}
for the neutralinos, and 
\begin{eqnarray}
\tilde w^\pm &=& \chi^\pm_1 + O({m_Z^2/\mu^{2}})\ ,\\
\tilde H_d^- &=& -\sqrt{2}\frac{c_\b c_W m_Z}{\mu}\chi^-_1 + \chi^-_2 +O({m_Z^2/\mu^{2}})\ ,\\
\tilde H_u^+ &=& -\sqrt{2}\frac{s_\b c_W m_Z}{\mu}\chi^+_1 + \chi^+_2+ O({m_Z^2/\mu^{2}})\ .
\end{eqnarray}
for the charginos.

In these bases, the mass eigenvalues are given by,
\begin{eqnarray}
m_{\chi^0_1} &=& M_2 - \frac{s_{2\b}c_W^2m_Z^2} {\mu} + O(m_Z^3/\mu^2)\ , \cr
m_{\chi^0_2} &=& M_1 - \frac{s_{2\b}s_W^2m_Z^2} {\mu} + O(m_Z^3/\mu^2)\ , \cr
m_{\chi^0_3} &=& \mu+ O(m_Z^3/\mu^2)\ , \cr
m_{\chi^0_4} &=& \mu+ O(m_Z^3/\mu^2)\ ,
\end{eqnarray}
and 
\begin{eqnarray}
m_{\chi_1^\pm} &=& M_2 - \frac{s_{2\b} c_W^2 m_Z^2}{\mu} + O(m_Z^3/\mu^2)\ ,\\
m_{\chi_2^\pm} &=& \mu + O(m_Z^3/\mu^2)\ .
\end{eqnarray}
In the followings, we call $\chi_1^0$ the neutral wino,
$\chi_2^0$ the bino, and $\chi_1^\pm$  the charged wino.

In terms of these mass eigenstates, 
the neutral wino-bino-Higgs couplings, i.e. $\chi^0_1$-$\chi^0_2$-$h$, 
are obtained from the gaugino-Higgs-higgsino interactions,
which lead to
\begin{eqnarray}
{\cal L}_{\chi^0_1-\chi^0_2-h} &=& \frac{g s_W s_{2\b} m_Z}{\mu} \, h\,\chi^0_1\,\chi^0_2 + c.c. \cr
&=& \frac{g s_W s_{2\b} m_Z}{\mu} \, h\,\bar{\Psi}_1^0\,\Psi^0_2 \ .
\end{eqnarray}
In the final expression, we have used the four component Majorana fermions.
The charged wino-bino-$W$-boson interactions are obtained 
from the gauge interactions of the wino leading to,
\begin{eqnarray}
{\cal L}_{\chi^\pm_1-\chi^0_2-W^\mp}  &=&g \frac{s_{2\b} s_{2 W} m_Z^2}{2(M_1 - M_2)\mu}
(
- \tilde \chi_1^{+\dagger} \sigma^\mu  \tilde b W^+_\mu
+\, \tilde \chi_1^{-\dagger} \sigma^\mu  \tilde b W^-_\mu \cr
&&
+\, \tilde b^{\dagger} \sigma^\mu  \tilde \chi_1^- W^+_\mu
- \, \tilde b^{\dagger} \sigma^\mu  \tilde \chi_1^+ W^-_\mu 
)\\
&=&-g \frac{s_{2\b} s_{2 W} m_Z^2}{2(M_1 - M_2)\mu}
(
\bar\Psi^{+}_1 \gamma^\mu W_\mu^+ \Psi^0_2
+\bar\Psi^{0}_2 \gamma^\mu W_\mu^- \Psi^+_1
)\ .
\end{eqnarray}

From these interactions, we obtain the decay widths of the bino,
As a result, we obtain the decay width,
\begin{eqnarray}
\Gamma_{\tilde b\to h + \tilde w} &=& \frac{1}{16\pi}\left(\frac{gm_Zs_{2\b} s_W}{\mu} \right)^2\, M_1
\left(1 + 2  \frac{M_2}{M_1}
+  \frac{M_2^2}{M_1^2}
-  \frac{m_h^2}{M_1^2}
\right)
\nonumber\\
&&\times \left(1-\frac{(M_2 + m_h)^2}{M_1^2} \right)^{1/2}
\left(1-\frac{(M_2 - m_h)^2}{M_1^2} \right)^{1/2} \ , \\
\G_{\tilde b\to W^\pm + \tilde w^\mp} &=& \frac{1}{16\pi}\left(\frac{gm_Zs_{2\b} s_W}{\mu} \right)^2\, M_1
\left(
1- \frac{M_2}{M_1}
\right)^{-2}
\nonumber\\
&&
\times \left( 
\left(
1 - \frac{M_2^2}{M_1^2} + \frac{m_W^2}{M_1^2}
\right)
\left(
1 - \frac{M_2^2}{M_1^2} - \frac{m_W^2}{M_1^2}
\right)
+ \frac{m_W^2}{M_1^2}
\left(
1 -6 \frac{M_2}{M_1}
+ \frac{M_2^2}{M_1^2}
 - \frac{m_W^2}{M_1^2}
\right)
\right)
\nonumber\\
&&\times \left(1-\frac{(M_2 + m_W)^2}{M_1^2} \right)^{1/2}
\left(1-\frac{(M_2 - m_W)^2}{M_1^2} \right)^{1/2} \ .
\end{eqnarray}
Notice that these decay widths coincides 
in the limit of $M_{1,2}\gg m_{W,h}$,
\begin{eqnarray}
\Gamma_{\tilde b\to h + \tilde w} =
\G_{\tilde b\to W^\pm + \tilde w^\mp}
=  \frac{1}{16\pi}\left(\frac{gm_Zs_{2\b} s_W}{\mu} \right)^2\, M_1
\left(1 +  \frac{M_2}{M_1}\right)^2
 \left(1-\frac{M_2^2 }{M_1^2} \right)\ ,
\end{eqnarray}
up to $O(m_{h,W}^2/M_1^2)$ corrections, which exemplifies 
the equivalence theorem.
With the equivalence theorem, in our mind, we simplify our analysis by fixing $Br_{\tilde b\to \tilde w^0} \simeq 1/3$.

Since the bino decay into the neutral wino is the two-body decay,
the energy distribution of the neutral wino is a flat distribution as shown 
in Fig.\,\ref{fig:dGdx} in between,
\begin{eqnarray}
E_{\tilde w}^{\rm max} &=& \frac{E_{\tilde b}}{m_{\tilde b}}E_{\tilde w}^{\rm CM} + 
 \frac{{ p}_{\tilde b}}{m_{\tilde b}}
{ p}_{\tilde w}^{\rm CM}  \ ,\\
E_{\tilde w}^{\rm min} &=& \frac{E_{\tilde b}}{m_{\tilde b}}E_{\tilde w}^{\rm CM} - 
 \frac{p_{\tilde b}}{m_{\tilde b}}
p_{\tilde w}^{\rm CM}\ ,
\end{eqnarray}
where $E_{\tilde w}^{\rm CM}$
and $p_{\tilde w}^{\rm CM}$
denote the energy and the size of the three-dimensional momentum
in the rest frame of the bino which are given by,
\begin{eqnarray}
E_{\tilde w}^{\rm CM} &=&\frac{m_{\tilde b}^2 + m_{\tilde w}^2 - m_{h}^2 }{2 m_{\tilde b}}\ ,\\
 p_{\tilde w}^{\rm CM} &=&\frac{\sqrt{(m_{\tilde b}^2 -( m_{\tilde w} + m_{h})^2)
(m_{\tilde b}^2 -( m_{\tilde w} - m_{h})^2)}}
{2 m_{\tilde b}}\ .
\end{eqnarray}
Similarly, the charged wino distribution is given by replacing the Higgs boson
masses with the $W$-boson mass.
In our actual analysis, we set $m_h = m_W = 0$\, which leads to
harder wino in the cascade decays of the gravitinos.
As we have discussed in section\,\ref{sec:SmallScale}, the harder the 
initial wino is, the more likely it is converted to the charged wino
which immediately loses its energy via the scattering processes 
with the thermal background.
Therefore, this assumption gives us conservative estimations
of possible imprints on the small-scale structure 
of the non-thermally produced wino dark matter.

\section{Elastic scattering of the neutral wino at the one-loop level}
\label{loopscatteing}

\begin{figure}[tb]
 \begin{center}
  \subfigure[]{
  \includegraphics[width=.3\linewidth]{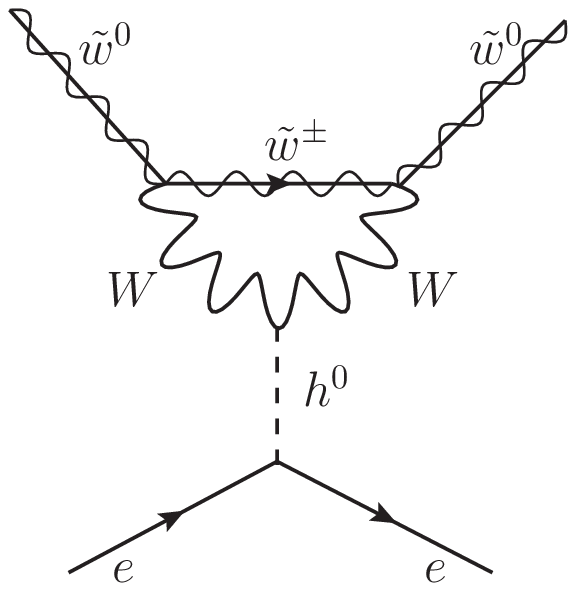}
  \label{fig:h_exchange}
  }
  \subfigure[]{
  \includegraphics[width=.3\linewidth]{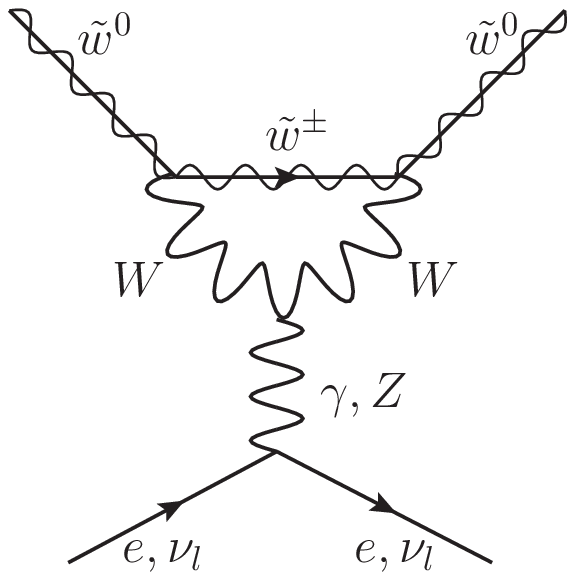}
   \label{fig:Z(gamma)_exchange_1}
  }
  \subfigure[]{
  \includegraphics[width=.3\linewidth]{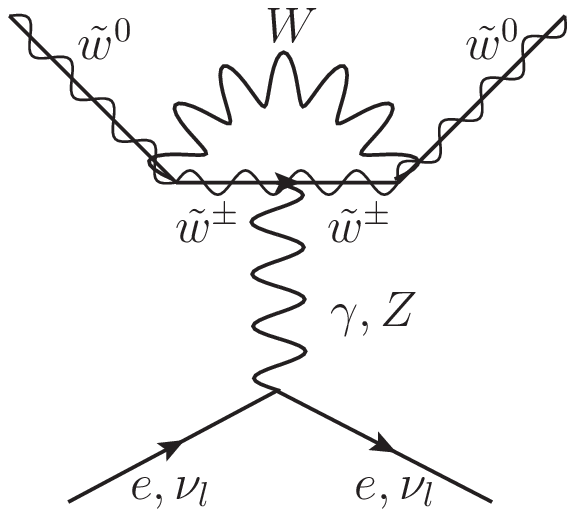}
  \label{fig:Z(gamma)_exchange_2}
  }
  \subfigure[]{
  \includegraphics[width=.3\linewidth]{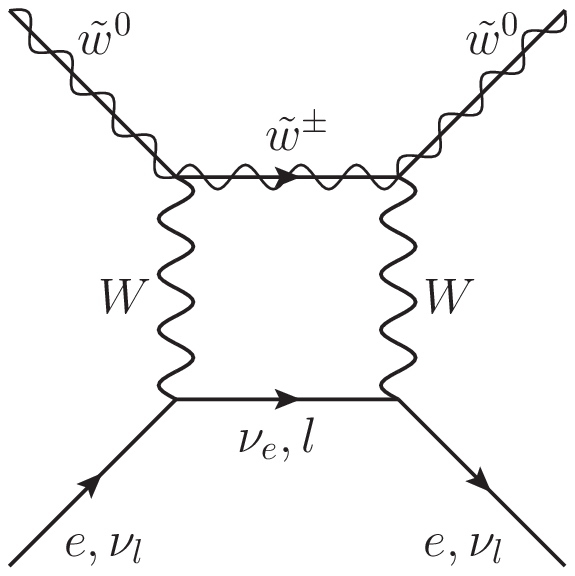}
  \label{fig:W_exchange}
  }
  \subfigure[]{
  \includegraphics[width=.3\linewidth]{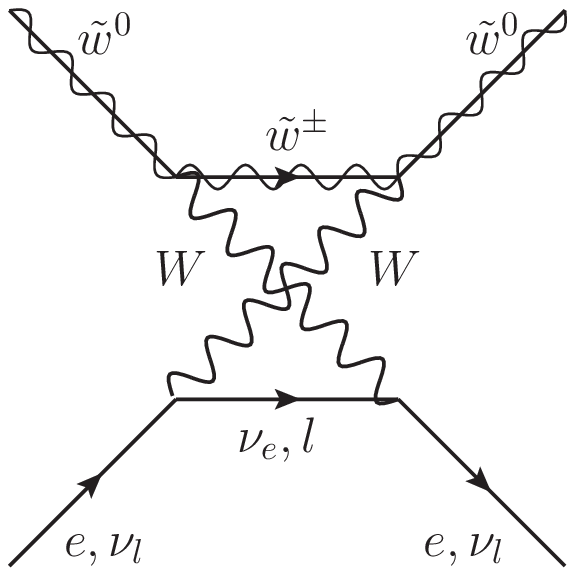}
  \label{fig:W_crossing}
  }
 \end{center}
\caption{\sl \small
The one-loop diagrams which contribute to the elastic scattering between the neutral wino and the thermal background in the decoupling limit of the sfermions, the higgsinos and the heavy Higgses.
}
\label{fig:loopdiagrams}
\end{figure}

In the decoupling limit of the sfermions, the higgsinos and the heavy Higgses,  the neutral wino doesn't have any tree level elastic interaction with the thermal background.
The one-loop diagrams shown in Fig.\,\ref{fig:loopdiagrams} contribute to the elastic scattering between the neutral wino and the thermal background. The contribution of the light Higgs exchange diagram (Fig.\,\ref{fig:h_exchange}) is negligible because the Yukawa coupling of the electron is small. 
The $\gamma,Z$ exchange diagrams (Fig.\,\ref{fig:Z(gamma)_exchange_1},\,\ref{fig:Z(gamma)_exchange_2}) originate from the one-loop correction to the ${\tilde w}^{0}-{\tilde w}^{0}-\gamma,Z$ vertex. The one-loop vertex correction consists of only vector-like interactions and vanishes by the charge-conjugation invariance. 

Thus, we should consider only box diagrams (Fig.\,\ref{fig:W_exchange},\,\ref{fig:W_crossing}).
In calculating the contribution of the box diagrams, we expand it by the incoming and the outgoing four-momenta of the thermal background particles in the loop. 
The incoming and the outgoing four-momenta of the thermal background particles are $O(E_{\tilde w} T / m_{\tilde w})$ at the rest frame of the neutral wino and much smaller than the wino mass $m_{\tilde w}$ and the weak boson mass $m_{W}$. 
This allows us to adopt the leading order of the expansion.
At the leading order, these diagrams yield an effective interaction,
\begin{eqnarray}
{\cal L}^{\rm eff}_{\rm int} =\sum_{f=e,\,\nu_{e},\,\nu_{\mu},\,\nu_{\tau}} \frac{1}{2} g_{\rm loop}^2 \! \left(\frac{m_{W}^2}{m_{\tilde w}^2} \right) G_{F}^2 m_{W}^2\,{\bar {\tilde w}^{0}}\gamma_{\mu}\gamma_{5}{\tilde w}^{0}\,{\bar f}\gamma^{\mu}P_{L}f
\end{eqnarray}
with
\begin{eqnarray}
g_{\rm loop} (x) =  \frac{1}{3\pi^2} \left( \frac{\sqrt{x}}{\sqrt{1-x/4}}(8-x-x^2) \arctan \left( \frac{2\sqrt{1-x/4}}{\sqrt{x}} \right) - x \left( 2 - (3 +x )\ln x \right) \right).
\end{eqnarray}
In the above expression, we have used four component majorana fermion for the neutral wino $\tilde w^{0}$.
Using the above effective interaction, we can find the reaction rate of the elastic scattering is given by,
\begin{eqnarray}
\Gamma_{{\tilde w}^{0}, \, {\rm elastic}}= \frac{135}{\pi^3} \zeta(5) g_{\rm loop}^2 \! \left(\frac{m_{W}^2}{m_{\tilde w}^2} \right) G_{F}^4 T^5 m_{W}^4 \frac{E_{{\tilde w}^{0}}^2}{m_{\tilde w}^2} \left( 1+\frac{p_{{\tilde w}^{0}}^2}{E_{{\tilde w}^{0}}^2} \right).
\end{eqnarray}


\end{document}